\documentclass[pdftex,superscriptaddress]{revtex4}

\usepackage{epstopdf, epsfig}
\usepackage{amssymb,amsmath,subfigure,mathrsfs,graphicx,enumerate,color,xcolor}

\newcommand{\dd}{\mathrm{d}}
\newcommand{\ee}{\mathrm{e}}
\def\d{{\rm d}}

\newcommand{\Fb}{\mathbf{F}}

\newcommand{\nb}{\mathbf{n}}

\newcommand{\ub}{\mathbf{u}}
\newcommand{\eb}{\mathbf{e}}

\newcommand{\Ub}{\mathbf{U}}\newcommand{\xb}{\mathbf{x}}\newcommand{\rb}{\mathbf{r}}

\let\grad\nabla
\let\grad\nabla

\newcommand{\pard}[2]{\frac{\partial #1}{\partial #2}}\newcommand{\totd}[2]{\frac{\mathrm{d}#1}{\mathrm{d}#2}}

\newcommand{\sm}[1]{{\color{black} #1}}

\begin{document}
\title{Viscous growth and rebound of a bubble near a rigid surface}
\author{S\'ebastien Michelin}
\email{sebastien.michelin@ladhyx.polytechnique.fr}
\affiliation{LadHyX -- D\'epartement de M\'ecanique, Ecole Polytechnique -- CNRS, 91128 Palaiseau, France}
\author{Giacomo Gallino}
\email{giacomo.gallino@epfl.ch}
\affiliation{Laboratory of Fluid Mechanics and Instabilities, EPFL, CH-1015 Lausanne, Switzerland}
\author{Fran\c cois Gallaire}
\email{francois.gallaire@epfl.ch}
\affiliation{Laboratory of Fluid Mechanics and Instabilities, EPFL, CH-1015 Lausanne, Switzerland}
\author{Eric Lauga}
\email{e.lauga@damtp.cam.ac.uk}
\affiliation{Department of Applied Mathematics and Theoretical Physics, University of Cambridge,\\  Cambridge CB3 0WA, United Kingdom}

\begin{abstract}

Motivated by the dynamics of microbubbles near catalytic surfaces in bubble-powered microrockets, we consider theoretically the growth of a free spherical bubble near a flat no-slip surface in a Stokes flow. The flow at the bubble surface is characterised by a constant slip length allowing to tune the hydrodynamic mobility of its surface and tackle in one formulation   both clean and contaminated bubbles as well as rigid shells. Starting with a bubble of infinitesimal size, the fluid flow and hydrodynamic forces on the growing bubble are obtained analytically. We demonstrate that, depending on the value of the bubble slip length relative to the initial distance to the wall, the bubble will either monotonically drain the fluid separating it from the wall, which will exponentially thin, or it will bounce  off the surface once before eventually draining the thin film.  
Clean bubbles are  shown to be a singular limit which always monotonically get repelled  from the surface. 
The   bouncing events for bubbles with finite slip lengths are    further analysed in detail in the lubrication limit. In particular, we identify the origin of  the reversal of the hydrodynamic force direction  as due to the change in the   flow pattern in the film between the bubble and the surface and to the associated lubrication pressure. Last, the final drainage dynamics of the film is observed to follow a universal algebraic scaling for all finite slip lengths.

\end{abstract}

\maketitle

\section{Introduction}

Gas bubbles can be found in a variety of   environments and applications, from  industrial~\citep{brennen1995} and geophysical~\citep{llewellin2005}   to biomedical systems~\citep{barak2005}. The growth and collapse of small bubbles involve complex combinations of mechanical and thermodynamic processes coupling  heat and mass exchanges to the fluid motion outside the bubble~\citep{plesset1977,prosperetti2017}. In the case of microscopic  bubbles, two main classes of problems have long been  of interest, namely  inertial~\citep{plesset1977,rayleigh1917} and diffusive bubble phenomena~\citep{prosperetti2017,epstein1950}. While the former focus on the dynamics of the fluid outside the bubble at short time scales and neglect physico-chemical exchanges between the fluid and the bubble, the latter consider specifically the forcing of the surrounding fluid by these exchanges and the associated surface dynamics on longer time scales.

The growth or collapse of an isolated bubble in an unbounded fluid  is an isotropic problem: the flow is purely radial and corresponds to a potential source singularity, regardless of the relative importance of viscous and inertial forces~\citep{rayleigh1917}. This simple symmetry is lost when bubble growth occurs under confinement where the presence of a nearby boundary restricts the fluid motion forced by the expansion of the bubble and generates a net force that results in a net migration of its centre. This process is particularly important for microscopic bubble-generating self-propelled colloids~\citep{wang2014} or micro-rockets~\citep{li2016}. In both cases, it is the growth of a gas bubble in the vicinity of the swimmer surface which leads to its   net displacement, which is all the more pronounced for highly-confined bubbles~\citep{li2014,gallino2018}. In fact, the formation of a bubble under asymmetric confinement is the essential ingredient for overcoming the viscous resistance of the surrounding fluid and achieving self-propulsion. In such systems, bubbles form due to the saturated concentration of a dissolved gas produced at the surface of a catalyst. Bubble nucleation often takes place  in the immediate vicinity of	 the confining surface where the gas concentrations are highest, although such nucleation is also known to involve a number of  intricate  physico-chemical processes~\citep{lv2017}. From a purely fluid mechanics standpoint, the flow generated by a growing bubble and resulting bubble motion may also be particularly relevant to boiling-type flows for which heat exchange and bubble generation occur in the immediate vicinity of the warmer surface, \sm{or in electrolysis where a locally-saturated gas concentration is obtained in the vicinity of an electrode}.

In the viscous regime (i.e.~for small bubbles or slow growth), a critical feature of the flow field generated by the confined bubble is found in the drainage of a thin lubricating film between the bubble and the wall. Expelling fluid from a thin layer can be prohibitively costly from a fluid mechanical point of view as it requires overcoming  viscous stresses that diverge for small film thicknesses~\citep{kimbook}. As a result, in the textbook example of \sm{a} sedimenting sphere under a constant net gravity force, the lubricating film thickness decreases exponentially in time and perfect contact with a flat surface is only achieved asymptotically (i.e.~for infinite time). In the case of a bubble growth, no such net force is applied to drag the bubble toward the wall and instead the only source of motion is the growth itself. Two fundamental questions then arise: (a) Does the lubricating film between a growing bubble and the nearby surface fully drain in finite or infinite time? (b) What are the associated scalings and asymptotic regimes for the film dynamics and resulting bubble motion?

The dynamics of a thin fluid film between two moving spheres, droplets or bubbles is a canonical example of the application of lubrication theory to compute  leading-order hydrodynamic forces by exploiting the near-parallel nature of the flow within the gap~\citep{kimbook,leal}. Classical examples include the axisymmetric approach of two rigid spheres, bubbles or droplets~\citep{barnocky1989} and the translation of a single such object near a rigid wall in the normal~\citep{cooley1969,hocking1973} or longitudinal direction~\citep{oneill1967}. In general, the motion of the fluid within the film is forced by the translation of the object in the direction normal or parallel to the wall under the effect of a constant force, 
e.g.~gravity in sedimentation problems~\citep{cooley1969,oneill1967}. The sedimentation of a rigid sphere (or bubble) toward a no-slip wall is hence known to lead to a contact in infinite time between the moving object and confining boundary as the  film drainage generates lubrication stresses that diverge with the inverse of the film thickness~\citep{leal}. The magnitude of the lubrication forces are known to depend both on the local curvature of the film's boundaries~\citep{kimbook} and their physical nature, e.g.~the presence of hydrodynamic slip at the surface of a bubble~\citep{hocking1973} or the viscosity ratio for a viscous droplet~\citep{barnocky1989}. 

In this paper we solve for the viscous dynamics of a growing gas bubble near a rigid surface. The bubble is  hydrodynamically-free and thus no external forcing is present, at least in the early times when buoyancy effects are negligible. Instead a relative kinematic forcing is imposed by the radial motion of the bubble's boundary, and the bubble itself may translate freely toward or away from the wall to balance the lubrication stresses. One purpose of our study is then to investigate the lubrication behaviour under this new category of forcing and its implication on the dynamics of the fluid within and outside the film and resulting bubble motion. 
A key ingredient to the generation of fluid motion is the nature of the local mechanical forcing resulting from the bubble growth, and fundamental differences can be found depending on the detailed properties of the bubble surface, specifically how it can sustain or apply shear stresses on the surrounding fluid. To provide a generic physical description of the growth-induced flows, a simple Navier-type slip length model is used to tune the nature of the surface flow, allowing to analyse  both ``clean'' bubble surfaces, which cannot sustain any shear stress (infinite slip length), and ``polluted'' or surfactant-laden surfaces, which may behave as the surface of rigid shells (finite or zero slip lengths). 

Furthermore, bubbles are intrinsically deformable objects and their shape is the result of  local balances between the inner uniform gas pressure, the outer spatially-dependent hydrodynamic stress and the effect of surface tension. For relatively-slow flows, the ratio of hydrodynamic stress to capillary pressure is  small, and allows us to restrict   our analysis to spherical bubbles, for which kinematics is entirely described by two parameters: the spatial position of their centre and their radius. In that case, the detailed evolution in time of the bubble radius does not have any influence on the geometric evolution of the fluid film and bubble position, which is solely determined by the radius of the bubble itself, hence the bubble radius can be substituted for time as a parameter for the problem (see also Ref.~\citep{gallino2018}).

We investigate the canonical problem of the growth of a single bubble in a semi-infinite fluid domain bounded by a flat rigid wall  as a function of the relative initial distance of the bubble (when its radius is infinitesimally small) compared to the  slip length of its boundary, assumed to remain constant. The problem is first addressed analytically for an arbitrary combination of bubble size, position and slip length in Section~\ref{sec:model} adapting the classical solution originally obtained by Stimson \& Jeffery~ \citep{stimson1926} for axisymmetric viscous flow in bi-spherical coordinates to the case of an inflating boundary (i.e.~for which a source of mass is effectively present in the bubble). This framework is used in Section~\ref{sec:growth} to characterise the dynamics of the bubble motion and film drainage, as well as a complete description of the associated flow field within the lubricating film and in the outer region. In Section~\ref{sec:forces}, the effect of bubble growth on the fluid gap and associated flow forces is studied by holding the film thickness constant in order to identify the effect of the slip length on the intensity (and direction) of the resulting lubricating force. This sheds some fundamental light on the strikingly different dynamics of bubbles with no-slip and stress-free surfaces, as a result of their differential ability to drive fluid out of the lubricating film. These observations are then rationalised   in the asymptotic limit of a thin film using lubrication theory in Section~\ref{sec:lub}, to obtain the dominant scalings for the hydrodynamic forces. This provides a complete characterisation of the different regimes for the bubble motion and fluid film dynamics in Section~\ref{sec:consequences}. Our results, as well as some perspectives, are finally discussed in Section~\ref{sec:conclusions}.

\section{Flow  and forces around a bubble near a wall}
\label{sec:model}
\subsection{Description of the problem}

\begin{figure}
\begin{center}
\includegraphics[width=.6\textwidth]{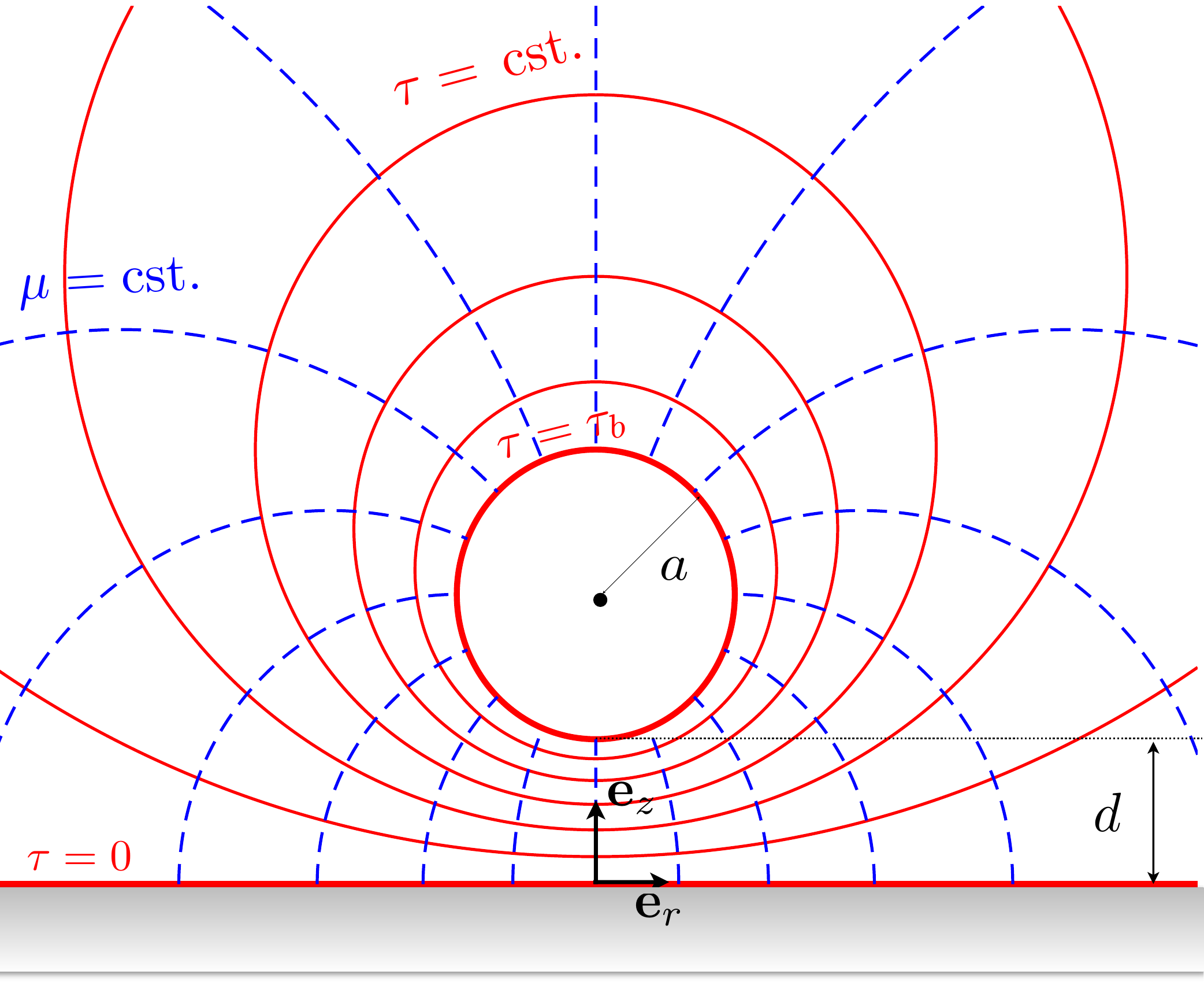}
\caption{Schematic of the problem along with the bi-spherical coordinate system. The surface of the bubble (resp.~the wall) corresponds to $\tau=\tau_b$ (resp.~$\tau=0$); $a$ and $d$ are the bubble's radius and minimum distance to the wall, respectively.}\label{fig:coord}
\end{center}
\end{figure}

We consider here the fluid motion and resulting displacement of a growing gas bubble initially located at a distance \sm{$d_0^*$} from a fixed flat infinite, rigid boundary $\partial \Omega_w$ with infinitesimal initial radius \sm{$a_0^*\ll d_0^*$}. At a given time $t$, the radius of the growing bubble is noted $a^*(t)$ and the shortest distance from the wall to its surface is noted \sm{$d^*(t)$} (see figure~\ref{fig:coord}). The fluid has mass density $\rho$ and dynamic viscosity $\eta$. The rate of change \sm{$\dot{a}^*(t)$}  of the bubble radius is prescribed here with a characteristic magnitude \sm{$U^*$}, and may result from gas dissolution within the bulk fluid or phase transition (e.g.~evaporation) at the liquid-gas interface.  \sm{Inertial} effects and bubble deformation are neglected (i.e.~\sm{$\mbox{Re}=U^*d^*_0/\nu\ll 1$ and $\mbox{Ca}=\eta U^*/\gamma\ll 1$}). \sm{Since buoyancy forces scale with the  volume of the bubble while hydrodynamic forces scale with its radius, buoyancy effects are also negligible for sufficiently  small  bubbles,  $a^* \ll \sqrt{\eta U^*/\rho g}$,  and the bubble in that case is hydrodynamically force-free.}

\sm{In the following, $U^*$ and $d_0^*$ are chosen as characteristic velocity and length scales, respectively, and the typical hydrodynamic pressure scales as $\eta U^*/d^*_0$. The non-dimensional kinematic variables are $a(t)=a^*(t)/d_0^*$ and $d(t)=d^*(t)/d_0^*$. The initial distance to the wall is now $d_0=1$.} The resulting viscous hydrodynamic problem for the flow velocity, $\ub$, and dynamic pressure, $p$, are solutions of the incompressible Stokes equations, written in non-dimensional form as
\begin{equation}
\nabla^2\ub=\grad p,\qquad \nabla\cdot\ub=0.
\end{equation} 
The impermeability condition at the surface of the bubble and no-slip condition at the wall take the form
\begin{equation}
\ub\cdot\nb=\Ub\cdot\nb+\dot{a}\quad\textrm{on   }\partial\Omega_b,\qquad \ub=0\quad\textrm{on   }\partial\Omega_w
\end{equation}
with $\nb$ the local unit normal vector pointing into the \sm{liquid} domain and $\Ub=(\dot{a}+\dot{d})\,\eb_z$ the translation velocity of the centre of the bubble and we use \sm{$\partial\Omega_b$ and $\partial\Omega_w$ to denote the surface of the bubble and the wall, respectively. }

To close the problem for $(\ub,p,\Ub)$, an additional condition must be applied on each boundary in order to describe the nature of the surface (e.g.~perfect slip, no-slip...). To account for impurity of the bubble and present a generic framework, we assume in the following a slip-length model
\begin{equation}
(\mathbf{I}-\nb\nb)\cdot(\ub-\ub_\textrm{ref})=\lambda(\mathbf{I}-\nb\nb)\cdot(\grad\ub+\grad\ub^T)\cdot\nb\quad\textrm{on   }\partial\Omega_b,\label{eq:dynbc}
\end{equation}
where $\lambda$ is the slip-length on the bubble, which characterises the purity of the interface: ${\lambda=\infty}$ corresponds to a perfect slip condition and a ``clean'' bubble, while $\lambda=0$ is equivalent to an inflating spherical shell (e.g.~armoured bubble with large surface concentration of surfactants). In what follows, it will also be referred to as the ``rigid shell'' limit: for an inflating sphere, this should be understood as a system for which the relative tangential velocity (i.e in the moving frame of its centre) is strictly zero while the relative normal velocity is given by $\dot{a}$. 
In the previous equation, $\ub_\textrm{ref}=\Ub+\dot{a}\nb$ is the reference velocity of the rigid shell from which the slip length model is defined. Finally, the flow velocity must satisfy the no-slip boundary condition, i.e.~$\ub=0$ on the wall, $\partial \Omega_w$.

Importantly, we assume throughout \sm{this work} that the value of the slip length $\lambda$ remains constant. If the slip length was used as model for the behaviour of the surface in the presence of surfactants, clearly one would expect the value of $\lambda$ to vary during the growth process, which would require a more detailed treatment of the surface boundary conditions. Yet, the presence of slip, despite being taken as constant  in this paper, allows us to treat a range of problems within a unique mathematical formulation and, as we will see below, to illustrate the singular nature of clean bubbles.

At each instant $t$ independently, the velocity and pressure fields are obtained by solving the hydrodynamic problem prescribed by the instantaneous geometry, $a(t)$ and $d(t)$, and the instantaneous translation velocity $\dot{d}(t)$ and inflation rate $\dot{a}(t)$. The total hydrodynamic force on the bubble is then computed by integrating the hydrodynamic stress on the bubble surface. Since the problem is  axisymmetric, we may write $\Fb=F\eb_z$ and using the linearity and instantaneous nature of the Stokes flow problem, it can be expressed as
\begin{equation}
F=\eb_z\cdot\int_{\partial\Omega_b}\boldsymbol\sigma\cdot\nb\,\dd S=C_T(\varepsilon,\tilde\lambda)a\dot{d}+C_I(\varepsilon,\tilde\lambda)a\dot{a},
\end{equation}
where $C_T$ and $C_I$ are the \sm{non-dimensional} force coefficients associated with the translation of the bubble (holding its radius $a$ constant) and the increase of its radius (holding the minimum bubble-wall distance $d$ constant), respectively, and which only depend on the relative magnitude of the three lengths characterizing the instantaneous problem, namely the gap width $d$, the bubble radius $a$ and the constant slip length $\lambda$ (i.e. the reference length scale $d_0$ associated with the initial condition is irrelevant here). Hence, we define them as functions of the \sm{instantaneous} relative gap width $\varepsilon=d/a$ and reduced slip length $\tilde\lambda=\lambda/a$. 

\subsection{Bi-spherical coordinates}
The viscous flow resulting from the motion of two spherical bodies, or of one sphere next to a wall, in an otherwise quiescent fluid can be obtained analytically in bi-spherical coordinates~\citep{stimson1926}, and has in fact been used extensively to obtain the forces and fluid dynamics around bubbles or rigid spheres~\citep{brenner1961,oneill1964,cooley1969,haber1973}. However, these previous studies consider specifically spheres of constant volume. The classical framework needs therefore to be adapted here to account for the growth of the bubble.

The geometry of a single spherical particle near a flat wall is conveniently described using bi-spherical coordinates $(\tau,\mu,\phi)$, obtained from the classical cylindrical polar coordinates $(r,\phi,z)$ associated with the axis of symmetry of the problem (Figure~\ref{fig:coord}), as
\begin{equation}\label{eq:bispherical}
r=\frac{k\sqrt{1-\mu^2}}{\cosh\tau-\mu},\qquad z=\frac{k\sinh\tau}{\cosh\tau-\mu},
\end{equation}
with $k$ a scaling constant defined from the problem's geometry, Eq.~\eqref{eq:geom_def}, and metric coefficients and unit vectors defined from $\pard{\xb}{\tau}=h_\tau\eb_\tau$, etc... as
\begin{align}
\eb_\tau&=\frac{1-\mu\cosh\tau}{\cosh\tau-\mu}\eb_z-\frac{\sinh\tau\sqrt{1-\mu^2}}{\cosh\tau-\mu}\eb_r,\qquad h_\tau=\frac{k}{\cosh\tau-\mu},\\
\eb_\mu&=\frac{\sinh\tau\sqrt{1-\mu^2}}{\cosh\tau-\mu}\eb_z+\frac{1-\mu\cosh\tau}{\cosh\tau-\mu}\eb_r,\qquad h_\mu=\frac{k}{(\cosh\tau-\mu)\sqrt{1-\mu^2}},\\
\eb_\phi&=-\cos\phi\,\eb_x+\sin\phi\,\eb_y,\qquad \qquad \qquad \qquad \,\,h_\phi=\frac{k\sqrt{1-\mu^2}}{\cosh\tau-\mu}\cdot
\end{align}
In what follows, we consider an axisymmetric problem so all the fields are independent of $\phi$. The surface of the bubble is defined by $\tau=\tau_b>0$ while the confining boundary corresponds to $\tau=\tau_w=0$. Hence, in Eq.~\eqref{eq:bispherical} the semi-infinite fluid domain corresponds to $(\tau,\mu)\in[\tau_w,\tau_b]\times[-1,1]$. Note that the configuration of a curved obstacle could be easily described using the same framework by taking $\tau_w\neq 0$.

The constants $k$ and $\tau_b$ are determined uniquely from the radius $a$ of the particle and the minimum distance $d$ to the wall, specifically
\begin{equation}\label{eq:geom_def}
k=\sqrt{d(d+2a)},\quad \cosh\tau_b=1+\frac{d}{a}\cdot
\end{equation}

\subsection{Viscous flow and forces on an inflating bubble near a confining boundary}

The general axisymmetric solution of Stokes equation outside an inflating bubble can be written as the sum of a potential and viscous contributions, $\ub=\ub^\textrm{pot}+\ub^\textrm{visc}$~\citep*{michelin2018},  where
\begin{equation}
\ub^\textrm{pot}=\grad\varphi, \qquad \varphi=-\frac{Q\,\ee^{\tau/2}(\cosh\tau-\mu)^{1/2}}{k\sqrt{2}},
\end{equation}
with $Q=a^2\dot{a}$ leads to a potential flow solution accounting for the net volume change of the bubble, and $\ub^\textrm{visc}$ is the classical axisymmetric volume-preserving solution obtained by Ref.~\citep{stimson1926} defined in terms of the axisymmetric streamfunction $\psi$ as
\begin{equation}\label{eq:gensol_volcons}
\psi=\frac{\chi(\tau,\mu)}{\Delta(\tau,\mu)^{3/2}},\quad \textrm{with   } \chi=\sum_{n=1}^\infty V_n(\mu)U_n(\tau)\quad\textrm{and  }\Delta(\tau,\mu)=\cosh\tau-\mu,
\end{equation}
and, noting $P_n(\mu)$ the Legendre polynomial of degree $n$,
\begin{align}
V_n(\mu)&=P_{n-1}(\mu)-P_{n+1}(\mu)=\frac{2n+1}{n(n+1)}(1-\mu^2)P_n'(\mu),\\
U_n(\tau)&=A_n\cosh\left(n-\frac{1}{2}\right)\tau+B_n\sinh\left(n-\frac{1}{2}\right)\tau+C_n\cosh\left(n+\frac{3}{2}\right)\tau+D_n\sinh\left(n+\frac{3}{2}\right)\tau.
\end{align}
The components of the velocity field and tangential stress can be obtained directly in terms of the streamfunction $\psi$ (or $\chi$) and velocity potential $\varphi$ as
\begin{align}
u_\tau=&-\frac{\Delta(\tau,\mu)^2}{k^2}\pard{\psi}{\mu}+\frac{\Delta(\tau,\mu)}{k}\pard{\varphi}{\tau}\label{eq:utau},\\
u_\mu=&\frac{\Delta(\tau,\mu)^2}{k^2\sqrt{1-\mu^2}}\pard{\psi}{\tau}+\frac{\Delta(\tau,\mu)\sqrt{1-\mu^2}}{k}\pard{\varphi}{\mu},\\
\sigma_{\tau\mu}=&\pard{}{\tau}\left(\frac{\Delta(\tau,\mu)u_\mu}{k}\right)+\sqrt{1-\mu^2}\pard{}{\mu}\left(\frac{\Delta(\tau,\mu)u_\tau}{k}\right)\nonumber\\
=&\frac{\Delta(\tau,\mu)^{3/2}}{k^3\sqrt{1-\mu^2}}\left[\pard{^2\chi}{\tau^2}-(1-\mu^2)\pard{^2\chi}{\mu^2}-\frac{3\chi}{4}\left(1+\frac{2\sinh^2\tau}{\Delta(\tau,\mu)^2}\right)\right]\nonumber\\
&+\frac{\sqrt{1-\mu^2}}{k^2}\left\{\pard{}{\tau}\left[\Delta(\tau,\mu)^2\pard{\varphi}{\mu}\right]+\pard{}{\mu}\left[\Delta(\tau,\mu)^2\pard{\varphi}{\tau}\right]\right\}.\label{eq:sigma_taumu}
\end{align}

The impermeability condition  on $\partial\Omega_b$ (where $\nb=-\eb_\tau$) and $\partial \Omega_w$ (where $\nb=\eb_\tau$) are written respectively as
\begin{equation}
u_\tau(\tau_b)=\ub\cdot\eb_\tau=(\dot{a}+\dot{d})\eb_z\cdot\eb_\tau-\dot{a},\qquad u_\tau(0)=0.
\end{equation}
These can be integrated along each boundary to obtain $\chi(\tau_b,\mu)$ and $\chi(0,\mu)$ (by symmetry, the vertical axis is always a streamline so that $\chi=0$ for $\mu=-1$) as
\begin{align}
\chi(0,\mu)&=\frac{Q(1-\mu)}{\sqrt{2}}\left[3\sqrt{\frac{1-\mu}{2}}-(2-\mu)\right]\label{eq:chi_wall},\\
\chi(\tau_b,\mu)&=-\frac{Q\ee^{\tau_b/2}}{\sqrt{2}}\left[\Delta(\tau_b,\mu)^2-\Delta(\tau_b,\mu)^{3/2}\Delta(\tau_b,-1)^{1/2}+\Delta(\tau_b,\mu)-\frac{\Delta(\tau_b,\mu)^{3/2}}{\Delta(\tau_b,-1)^{1/2}}\right]\nonumber\\
&+Q\sinh^2\tau_b\left[\Delta(\tau_b,\mu)^{1/2}-\frac{\Delta(\tau_b,\mu)^{3/2}}{\Delta(\tau_b,-1)}-\frac{1-\mu^2}{2\Delta(\tau_b,\mu)^{1/2}}\right]-\frac{k^2\dot{d}(1-\mu^2)}{2\Delta(\tau_b,\mu)^{1/2}},\label{eq:chi_bubble}
\end{align}
with $\Delta(\tau,\mu)$ defined in Eq.~\eqref{eq:gensol_volcons}. Projecting Eqs.~\eqref{eq:chi_wall} and \eqref{eq:chi_bubble} onto $P'_n(\mu)$ leads to 
\begin{align}
U_n(0)=&\frac{3Q\sqrt{2}}{(2n-1)(2n+1)(2n+3)}-\frac{\delta_{n1}Q\sqrt{2}}{3},\label{eq:Un1}\\
U_n(\tau_b)=&-\frac{3 Q\sqrt{2}}{4(2n+1)}\left[\frac{\ee^{-(n+\frac{3}{2})|\tau_b|}}{2n+3}-\frac{\ee^{-(n-\frac{1}{2})|\tau_b|}}{2n-1}\right]\left(1+2\sinh^2\frac{\tau_b}{2}-\frac{2n(n+1)\sinh^2\tau_b}{3}\right)\nonumber\\
&+ \frac{Q\sqrt{2}\sinh^2\tau_b}{2(2n+1)}\ee^{-(n+\frac{1}{2})|\tau_b|}-\frac{\delta_{n1}Q\ee^{\tau_b/2}\sqrt{2}}{3}-\frac{k^2 n(n+1)\sqrt{2}}{2(2n+1)}\left[\frac{\ee^{-(n-\frac{1}{2})|\tau_b|}}{2n-1}-\frac{\ee^{-(n+\frac{3}{2})|\tau_b|}}{2n+3}\right]\dot{d}.\label{eq:Un2}
\end{align}

The no-slip condition at the wall and slip condition on the surface of \sm{the} bubble are respectively given by 
\begin{equation}
u_\mu(0,\mu)=0,\qquad -\lambda\sigma_{\tau\mu}(\tau_b,\mu)=u_\mu(\tau_b,\mu)-(\dot{a}+\dot{d})\, \eb_\mu\cdot\eb_z.
\end{equation}
which can be rewritten respectively as
\begin{align}
\sum_{m=1}^\infty V_m(\mu)U_m'(0)=&-\frac{Q}{2\sqrt{2}}(1-\mu^2),
\end{align}
and
\begin{align}
\sum_{m=1}^\infty V_m(\mu)\Bigg\{-\lambda\Bigg[U_m''(\tau_b)+&\left(m-\frac{1}{2}\right)\left(m+\frac{3}{2}\right)U_m(\tau_b)\Bigg]\Delta(\tau_b,\mu)^2-k\Delta(\tau_b,\mu)U_m'(\tau_b)
+\frac{3\sinh\tau_b}{2}(k+\lambda\sinh\tau_b)U_m(\tau_b)\Bigg\}\nonumber\\&=\frac{Q\ee^{\tau_b/2}}{2\sqrt{2}}(1-\mu^2)\Delta(\tau_b,\mu)\Big[3\lambda(\ee^{-\tau_b}-\mu)+k\Big]-\frac{k\sinh\tau_b(1-\mu^2)}{\Delta(\tau_b,\mu)^{1/2}}\left(Q\sinh^2\tau_b+k^2\dot{d}\right).
\end{align}
 Projecting these two equations along $P_n'(\mu)$, and using Eqs.~\eqref{eq:utau}--\eqref{eq:sigma_taumu} leads to
\begin{align}
&U_1'(0)=-\frac{Q}{3\sqrt{2}},\qquad U_{n\geq 2}'(0)=0,\label{eq:Unddot1}
\end{align}
and
\begin{align}
-\lambda&\left[\cosh^2\tau_b+\frac{2n^2+2n-3}{(2n-1)(2n+3)}\right]\tilde U_n''(\tau_b)+2\lambda\cosh\tau_b\left[\frac{(n+1)\tilde U_{n-1}''(\tau_b)+n\tilde{U}_{n+1}''(\tau_b)}{2n+1}\right]\nonumber\\
&-\frac{\lambda n(n+1)}{2n+1}\left[\frac{\tilde U_{n+2}''(\tau_b)}{2n+3}+\frac{\tilde U_{n-2}''(\tau_b)}{2n-1}\right]+\frac{(n+1)kU_{n-1}'(\tau_b)+nkU_{n+1}'(\tau_b)}{2n+1}\nonumber\\
&-k\cosh\tau_b U_n'(\tau_b)+\frac{3\sinh\tau_b}{2}(\lambda\sinh\tau_b+1)U_n(\tau_b)=f_n(\mu),\label{eq:Unddot2}
\end{align}
with
\begin{align}
f_n(\mu)=-k\sinh\tau_b&\left(\frac{n(n+1)\sqrt{2}}{2n+1}\right)\left[\frac{\ee^{-(n-1/2)|\tau_b|}}{2n-1}-\frac{\ee^{-(n+3/2)|\tau_b|}}{2n+3}\right](k^2\dot{d}+Q\sinh^2\tau_b)\nonumber\\
+\frac{Q\ee^{\tau_b/2}}{2\sqrt{2}}&\left[(k+3\lambda\sinh\tau_b)\left(\frac{2}{3}\cosh\tau_b\delta_{n,1}-\frac{2}{5}\delta_{n,2}\right)+3\lambda\left(\frac{10\cosh^2\tau_b+2}{15}\delta_{n,1}-\frac{4}{5}\cosh\tau_b\delta_{n,2}+\frac{8}{35}\delta_{n,3}\right)\right],\\
\tilde U_n''(\tau_b)=U_n''(\tau_b)+&(n-\frac{1}{2})(n+\frac{3}{2})U_n(\tau_b).
\end{align}
Truncating the expansion of the streamfunction $\psi$, Eq.~\eqref{eq:gensol_volcons}, to the first $N$ modes, \eqref{eq:Un1}, \eqref{eq:Un2}, \eqref{eq:Unddot1} and \eqref{eq:Unddot2} \sm{provide} a set of $4N\times 4N$ linear equations for the coefficients $A_n$, $B_n$, $C_n$ and $D_n$ defining $U_n(\tau)$ and the flow around the bubble uniquely as a function of $Q=a^2\dot{a}$ and $\dot{d}$. \sm{In the following, $N$ was chosen sufficiently large so as to ensure that truncation errors are negligible: in practice, $N$ ranged from 100 to 1 600 depending on the aspect ratio $d/a$ (thinner gaps require higher azimuthal resolution).} 

\sm{The potential contribution $\ub^\textrm{pot}$ associated with the bubble growth does not contribute to the net force on the bubble \citep{michelin2018}. The hydrodynamic force on the bubble therefore takes the same form as} the classical result of Ref.~\citep{stimson1926} 
\begin{equation}
F=\frac{2\pi\sqrt{2}}{k}\sum_{n=1}^\infty(2n+1)(A_n+B_n+C_n+D_n)=C_T(\varepsilon,\tilde\lambda)a\dot{d}+C_I(\varepsilon,\tilde\lambda)a\dot{a}.\label{eq:force}
\end{equation}
\sm{It should however be noted that the values of the coefficients $A_n$, $B_n$, $C_n$ and $D_n$ differ from those obtained by Ref.~{stimson1926} as they are related to the bubble growth problem through the boundary conditions. Equation~\eqref{eq:force}} defines the force coefficients $C_T(\varepsilon,\tilde\lambda)$ and $C_I(\varepsilon,\tilde\lambda)$ uniquely.  For a force-free bubble ($F=0$ for all time), this equation provides a direct relationship between the growth rate of the bubble $\dot{a}$ and the rate of change of its smallest distance to the confining wall $\dot{d}$ (or alternatively the translation velocity of its centre of mass $\dot{a}+\dot{d}$).

 \section{Growth of a force-free bubble near a surface}
\label{sec:growth}
We now turn to the main questions at the heart of this work, namely characterising (i) the motion of an inflating bubble (or more generally an inflating body) in proximity to a confining rigid surface in Stokes flow and (ii) the associated flow within the fluid gap in order to determine the extent to which  growth results in a complete drainage of the fluid film.

\subsection{Dynamic evolution of the fluid gap width}
The growth of a single force-free bubble is thus considered. At $t=0$ the bubble has an infinitesimal radius, i.e.~$a(0)=a_0=0$, and is located at a distance $d(0)=d_0=1$ from the planar rigid wall (this initial distance was chosen here as a reference length scale). For $t>0$, the bubble radius grows (e.g.~due to dissolved gas absorption or evaporation) and the bubble translates under the effect of the resulting flow. Neglecting any external force  (e.g.~buoyancy) on the bubble other than hydrodynamic stresses, the evolution of the fluid gap thickness $d(t)$ follows  from the previous analysis by writing $F=0 $ as
\begin{equation}
\totd{d}{t}=-\frac{C_I(\varepsilon,\tilde\lambda)}{C_T(\varepsilon,\tilde\lambda)}\totd{a}{t}\cdot
\end{equation}
In the previous equation, $\varepsilon(t)\equiv d(t)/a(t)$ and $\tilde\lambda(t)\equiv \lambda/a(t)$ denote the {instantaneous} relative gap width and relative slip length, respectively. 

Equivalently, \sm{the distance to the wall can be found as a function of the bubble radius only $d\equiv D(a)$} (for fixed initial distance $d_0$ and slip-length $\lambda$) and \sm{$D(a)$ satisfies}
\begin{equation}
\totd{D}{a}=-\frac{C_I(D/a,\lambda/a)}{C_T(D/a,\lambda/a)},\qquad D(0)=d_0.\label{eq:gap_dyn}
\end{equation}
The linearity and time-independence of the steady Stokes equations essentially transform the problem into a purely geometric one, providing a direct link between $a$ and $d$. Physico-chemical details determining the growth rate of the bubble will only affect the rate at which this parametric trajectory $d=D(a)$ is followed. 

\begin{figure}
\begin{center}
\begin{tabular}{cc}
\subfigure[No slip ($\lambda=0$)]{\includegraphics[width=.45\textwidth]{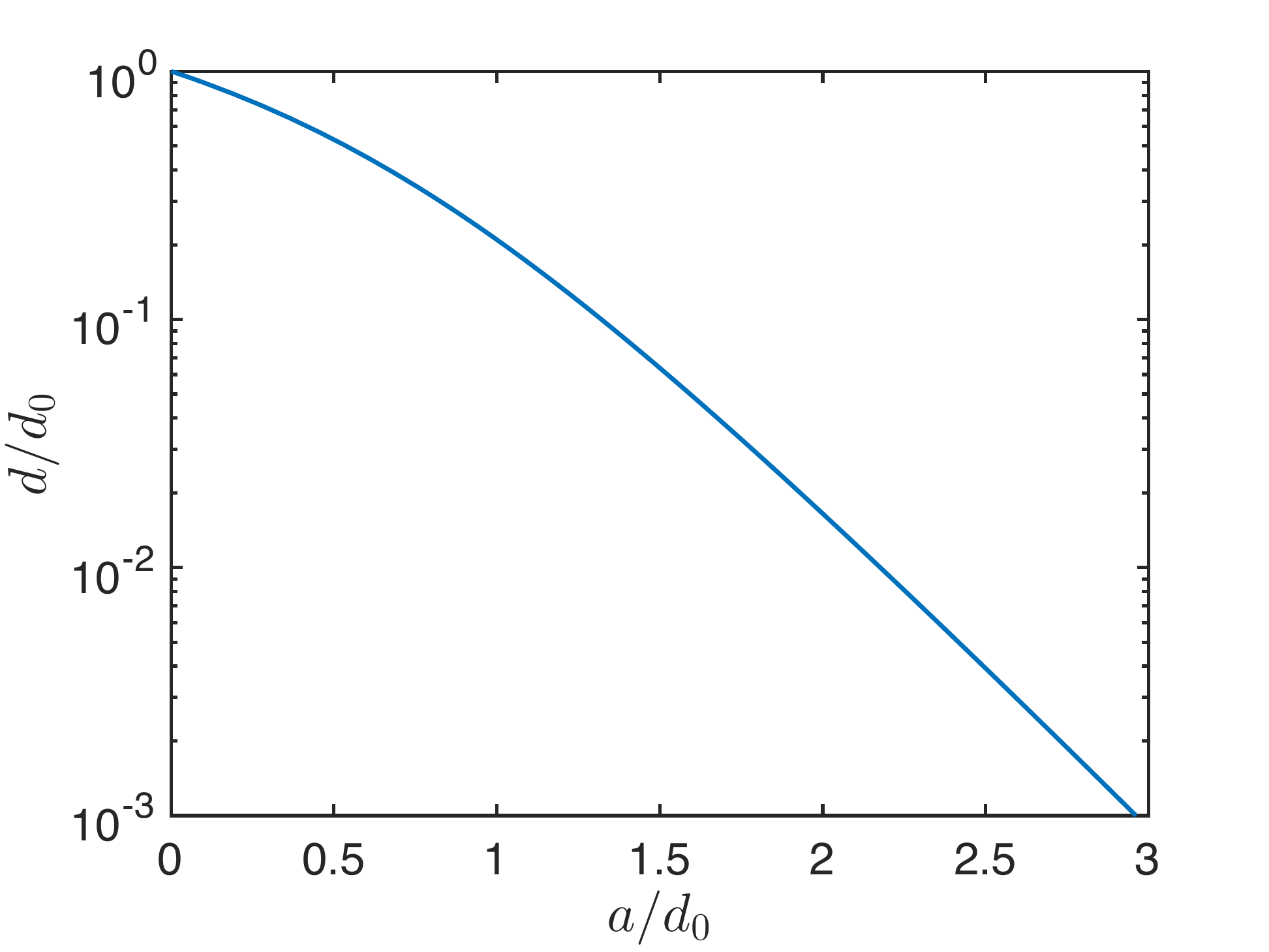}}&
\subfigure[Perfect slip ($\lambda=\infty$)]{\includegraphics[width=.45\textwidth]{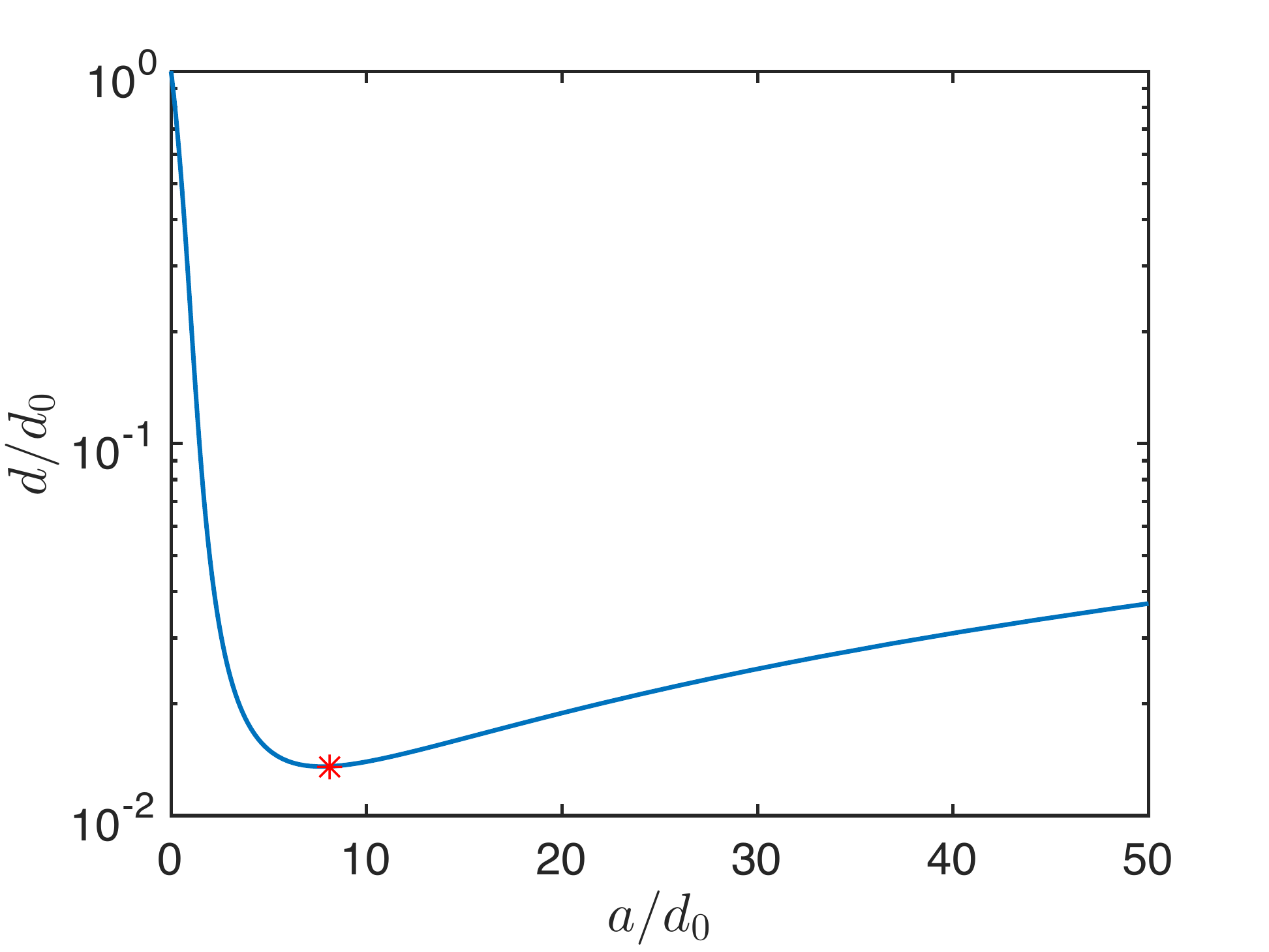}}\\
\multicolumn{2}{c}{\subfigure[General case]{\includegraphics[width=.7\textwidth]{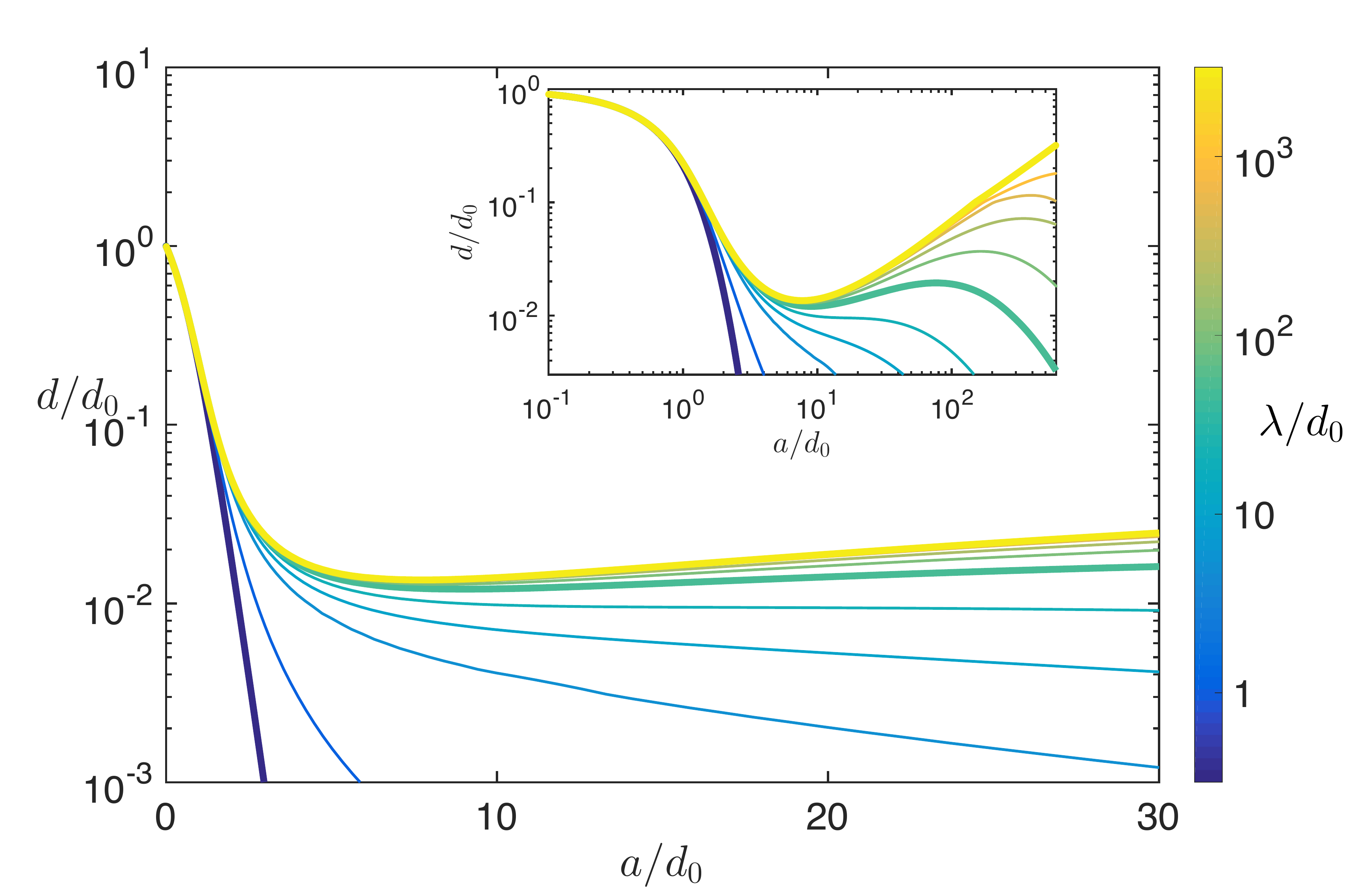}}}
\end{tabular}
\caption{Evolution of the dimensionless gap width, $d/d_0$, between a growing rigid shell (a, $\lambda=0$) or clean bubble (b, $\lambda=\infty$) and a no-slip wall for increasing radius $a$. In panel (b), the red star denotes the minimum gap width $d_\textrm{min}/d_0\approx 1.35\%$ attained for $a/d_0\approx 8.1$ when $\lambda=\infty$. (c) Evolution of the gap width for varying $\lambda/d_0$. Results are shown for $\lambda/d_0=0$, $1$, $5$, $10$, $20$, $50$, $100$, $200$, $500$, $1000$ and $\infty$, and the three thicker lines correspond to the three configurations chosen to analyse the flow field in figure~\ref{fig:flow}. Inset:  results shown  for a longer range of $a/d_0$ values.}
\label{fig:growth_bubble}
\end{center}
\end{figure}

The evolution of $d$ with the bubble radius $a$ is shown on figure~\ref{fig:growth_bubble} for different values of the slip length $\lambda$ (measured in units of $d_0$). The early dynamics is identical in all cases when the bubble is still far from the wall in comparison with its radius (i.e.~$d\gg a$). In that case, the problem is  equivalent to an isolated bubble in unbounded flow, whose growth does not lead to any translation, the flow is isotropic, $\ub=(a^2\dot{a}/r^3)\rb$, for any $\lambda$, and by geometric construction $\dot{d}=-\dot{a}$.

When the effect of confinement is significant (i.e.~$d\lesssim a$), a fundamentally-different behaviour is observed depending on the nature of the bubble boundary via the value of $\lambda/d_0$. In the case of a rigid spherical shell ($\lambda=0$, no tangential slip), the gap between the sphere and the wall decreases monotonically during the growth process (Figure~\ref{fig:growth_bubble}a), and the asymptotic dynamics is an exponential decay of the thin lubricating film, leading to its complete drainage in infinite time (i.e.~$d\rightarrow 0$ when $a\rightarrow\infty$). In practice, surface roughness effects will lead to contact with the confining wall in finite time when the thickness of the film is reduced below the typical roughness scale. 

A clean bubble, whose surface cannot sustain any tangential stress ($\lambda=\infty$), presents a strikingly different dynamics. After the early bubble growth and closing of the lubricating film, confinement effects slow down and even reverse the fluid drainage within the film when $\varepsilon=d/a$ is small enough. A minimum value $d_m/d_0\approx 0.014$ of the fluid gap is reached for $a_m/d_0\approx 8.1$, beyond which any further increase in bubble size  leads to a reopening of the fluid gap (i.e.~increasing $d(t)$, figure~\ref{fig:growth_bubble}b). Near the axis of symmetry of the problem the net flow must therefore now be oriented inward.

For intermediate slip lengths ($0<\lambda<\infty$), two different behaviours are observed:
\begin{itemize}\renewcommand{\labelitemi}{$-$}
\item[(I)]{When $\lambda/d_0\lesssim 20$, a monotonic drainage of the film  is found, similarly to, albeit slower than, the strictly rigid limit $\lambda=0$;}
\item[(II)]{When $\lambda/d_0\gtrsim 20$, a rebound of the film thickness is observed at intermediate times  in response  to confinement. After an initial decrease of $d(t)$ with increasing bubble size,  further increase of the bubble radius $a(t)$ leads to a repulsion of the bubble by the wall and an increase in the thickness of the fluid layer. This increase is however only transient, and eventually drainage is observed when $a(t)/d_0$ is large enough. The ``duration'' of this transitory repulsion (in terms of the range of $a(t)$) is an increasing function of the value of $\lambda/d_0$, i.e.~of the inability of the bubble surface to sustain shear stresses.}
\end{itemize}

\begin{figure}
\begin{center}
\includegraphics[width=.85\textwidth]{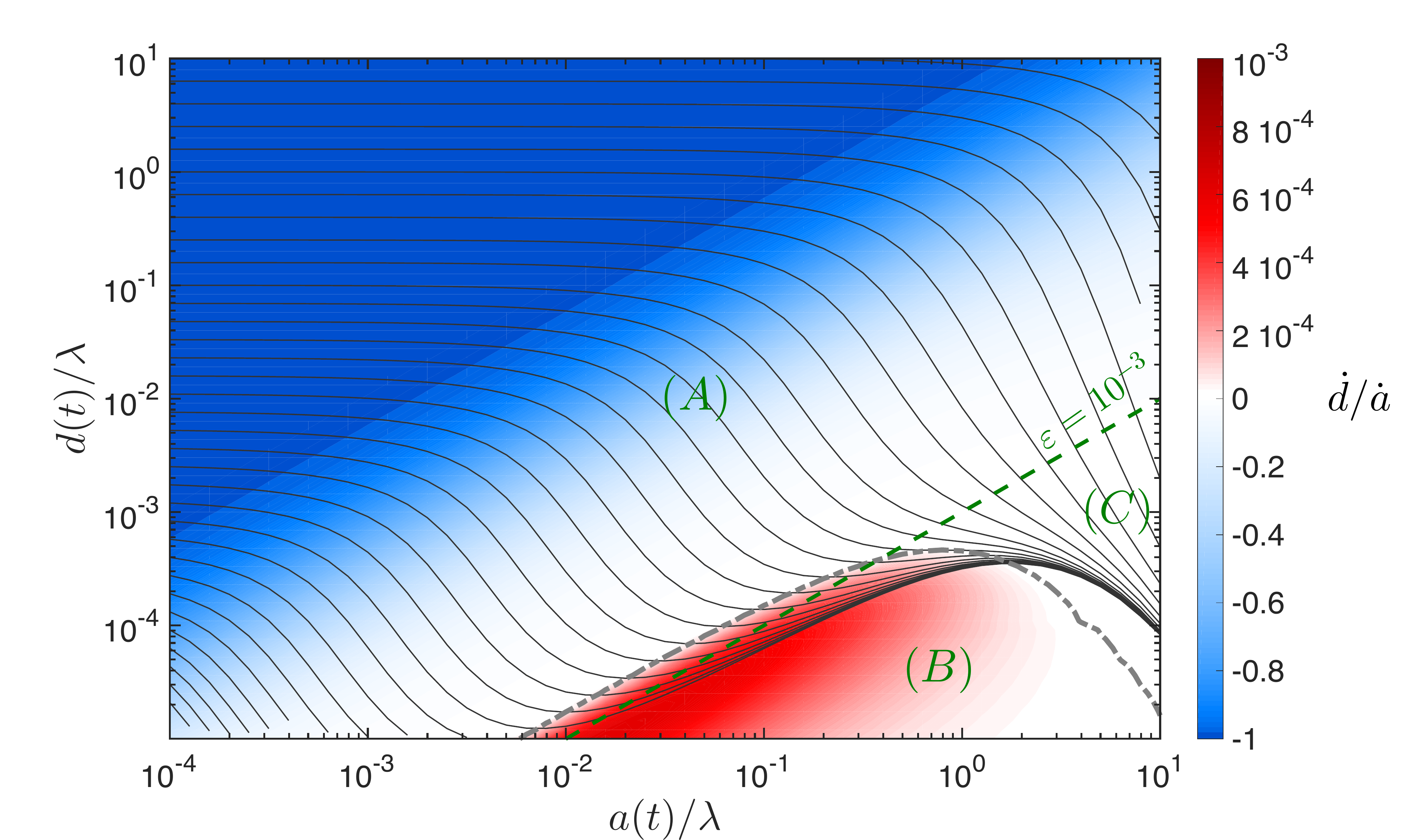}
\caption{Phase-space trajectory ($a(t)/\lambda$, $d(t)/\lambda$) for different values of the initial bubble-to-wall distance $d_0$ (with zero initial radius, $a_0=0$) relative to the slip length $\lambda$ (solid lines). The rate of change of the gap thickness ($\d D/\d a=\dot{d}/\dot{a}$) is also shown (color) with the dash-dotted grey line marking the boundary between collapsing and expanding lubrication film. Note the difference of scales for positive and negative velocity. A line of constant relative gap width ($\varepsilon=d/a=10^{-3}$) is also shown for reference (green dashed). The three different dynamical regimes (A), (B) and (C) as well as their boundaries are also shown (see text).}\label{fig:growth_general}
\end{center}
\end{figure}

These results can be alternatively presented by holding the material properties of the problem fixed, i.e.~representing $d/\lambda$ as a function of $a/\lambda$, with results shown in figure~\ref{fig:growth_general}. Three different regions can be distinguished:
\begin{itemize}\renewcommand{\labelitemi}{$-$}
\item[(A)]{$\varepsilon=d/a\gg 10^{-3}$: this corresponds to the regime of    {low confinement}. As the bubble grows, the fluid layer is drained monotonically. For $\varepsilon\gtrsim 10^{-1}$, this drainage is approximately linear, as discussed previously. The effect of $\lambda$ is only marginal and limited to the smaller values of $\varepsilon$ when wall influence induces non-isotropic flow.}
\item[(B)]{$\varepsilon=d/a\lesssim 10^{-3}$ and $a\ll \lambda$: in this lubrication regime, the drainage of the fluid layer is reversed (i.e.~positive values of $\dot{d}$ or $\d D/\d a$) by the wall-induced confinement, and this corresponds to a  {bubble repulsion} by the wall. In that region, trajectories follow approximately lines of constant $\varepsilon$ and the growth of the fluid layer is slightly slower than that of the radius of the bubble.}
\item[(C)]{$\varepsilon=d/a\lesssim 10^{-3}$ and $a\gg \lambda$: in this lubrication regime, monotonic film drainage is observed leading to a contact in infinite time between the bubble surface and the confining wall.}
\end{itemize}

The boundary between (B) and (C) is described qualitatively above by comparing the radius of the bubble to the slip length on the bubble and we will characterise this  boundary   more quantitatively in Section~\ref{sec:consequences}. For all inflating bubbles, phase-space trajectories, $D(a)$, always initiate in region (A) and eventually end in region (C) with the final drainage of the film (and contact between the bubble and the wall) for $a=\infty$. Depending on the initial distance relative to $\lambda$ (i.e.~$d_0/\lambda$), phase-space trajectories may or may not cross region (B), corresponding to the two different behaviours identified on figure~\ref{fig:growth_bubble}, namely a monotonic decrease of the film layer thickness (small $\lambda/d_0$, case I) or a transitory rebound (large $\lambda/d_0$, case II).

Importantly, for any $\lambda\neq\infty$, the lubrication film will therefore always asymptotically drain in infinite time. In the case of a rebound of the film thickness, how long this rebound lasts (i.e.~how large a radius the bubble needs to achieve for bubble growth to lead to a decrease in the gap thickness) scales linearly with $\lambda$ (on figure~\ref{fig:growth_general}, this corresponds to the change in direction of the phase space trajectories for $a\approx 2\lambda$).

A surprising result can also be observed on figure~\ref{fig:growth_general}, namely the existence of a region of the physical space (for $d/\lambda\leq 10^{-3}$ and intermediate $a/\lambda$) through which no trajectory originated at $a=0$ can be found. It  is thus not possible to create bubbles with $(a,d)$ within this region, starting from an initially infinitesimal bubble without exerting an external force.

In practical situations, the mathematical limit $d\rightarrow 0$ is not necessarily relevant as finite roughness effects may lead to the contact between the bubble surface and the wall at a finite value of $D$. This is  equivalent to introducing a finite lower limit for $d$ at which the evolution described by this model becomes invalid. \sm{Then, two families} of bubbles can be identified:  bubbles  initially close to the wall (small $d_0/\lambda$), for which the effective contact to the wall is made rather quickly and for small bubble radius, and  bubbles which start further away from the wall, and first experience a rebound in gap width before touching the wall at much larger values of $a$.

\subsection{Flow field within the gap}
\begin{figure}
\begin{center}
\begin{tabular}{ccc}
(a) Rigid shell ($\lambda/d_0=0$) & (b) $\lambda/d_0=50$ & (c) Clean bubble ($\lambda/d_0=\infty$) \\
\includegraphics[width=.33\textwidth]{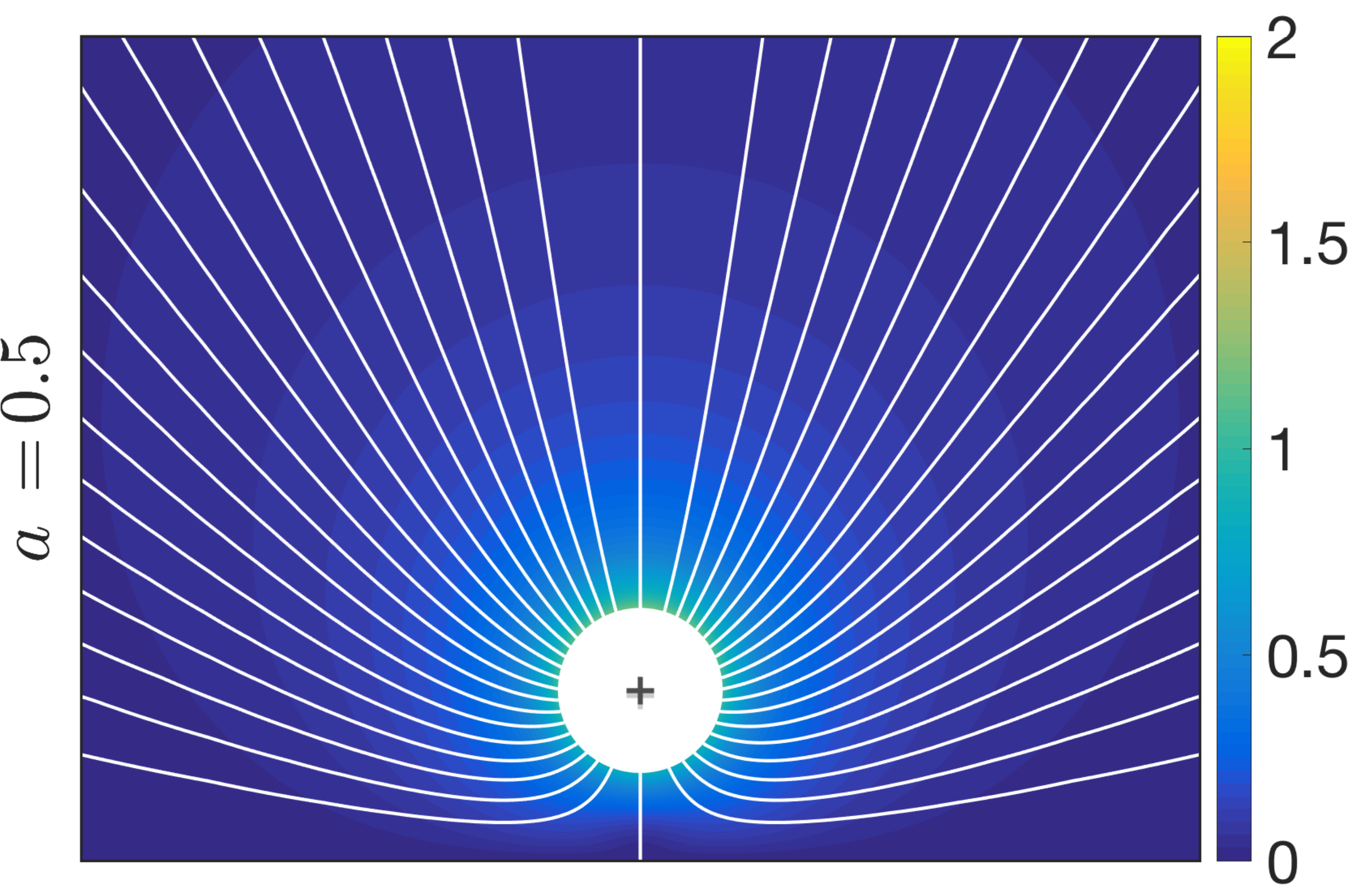} &
\includegraphics[width=.33\textwidth]{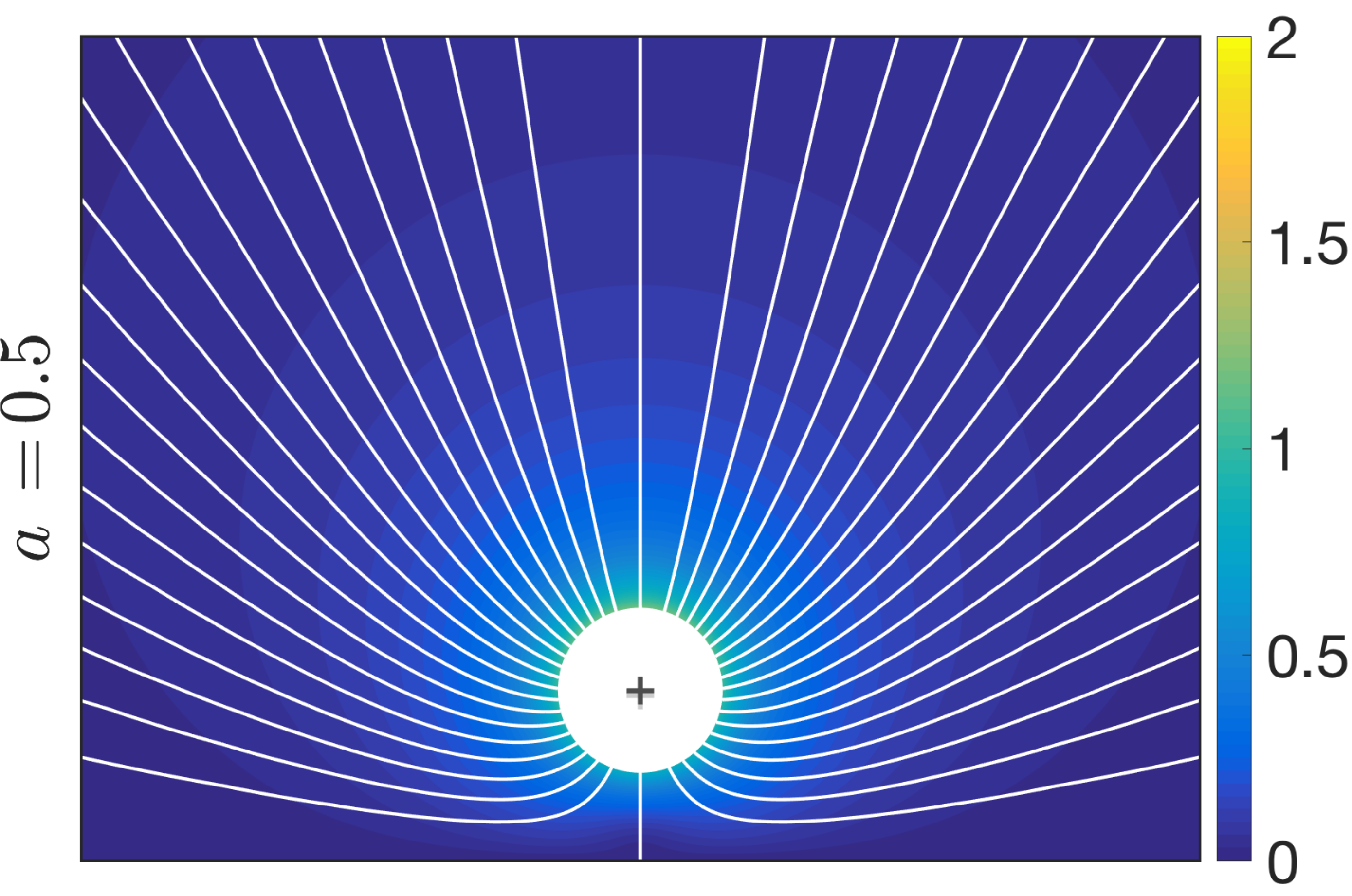} &
\includegraphics[width=.33\textwidth]{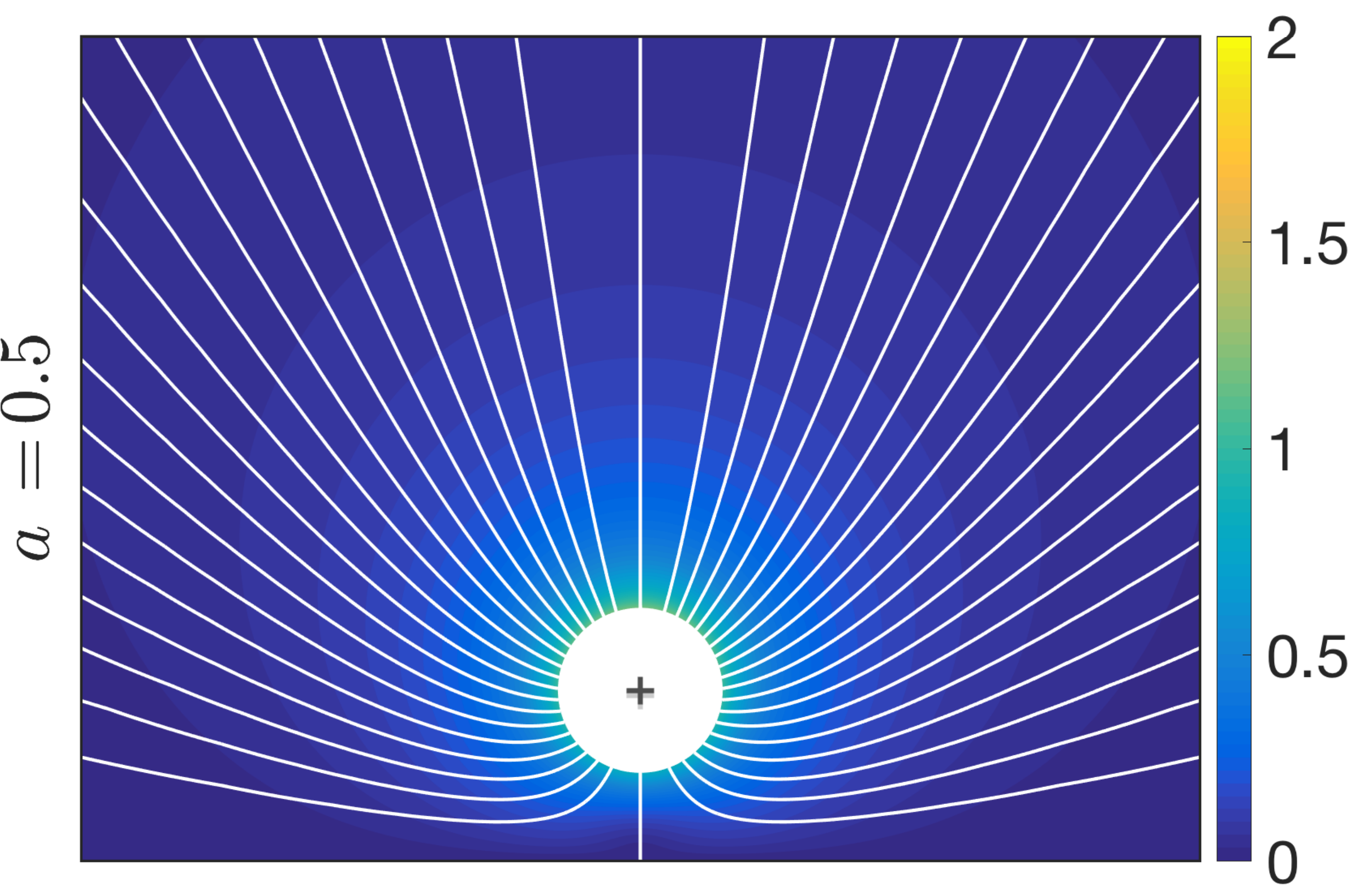}\\
&&\\
\includegraphics[width=.33\textwidth]{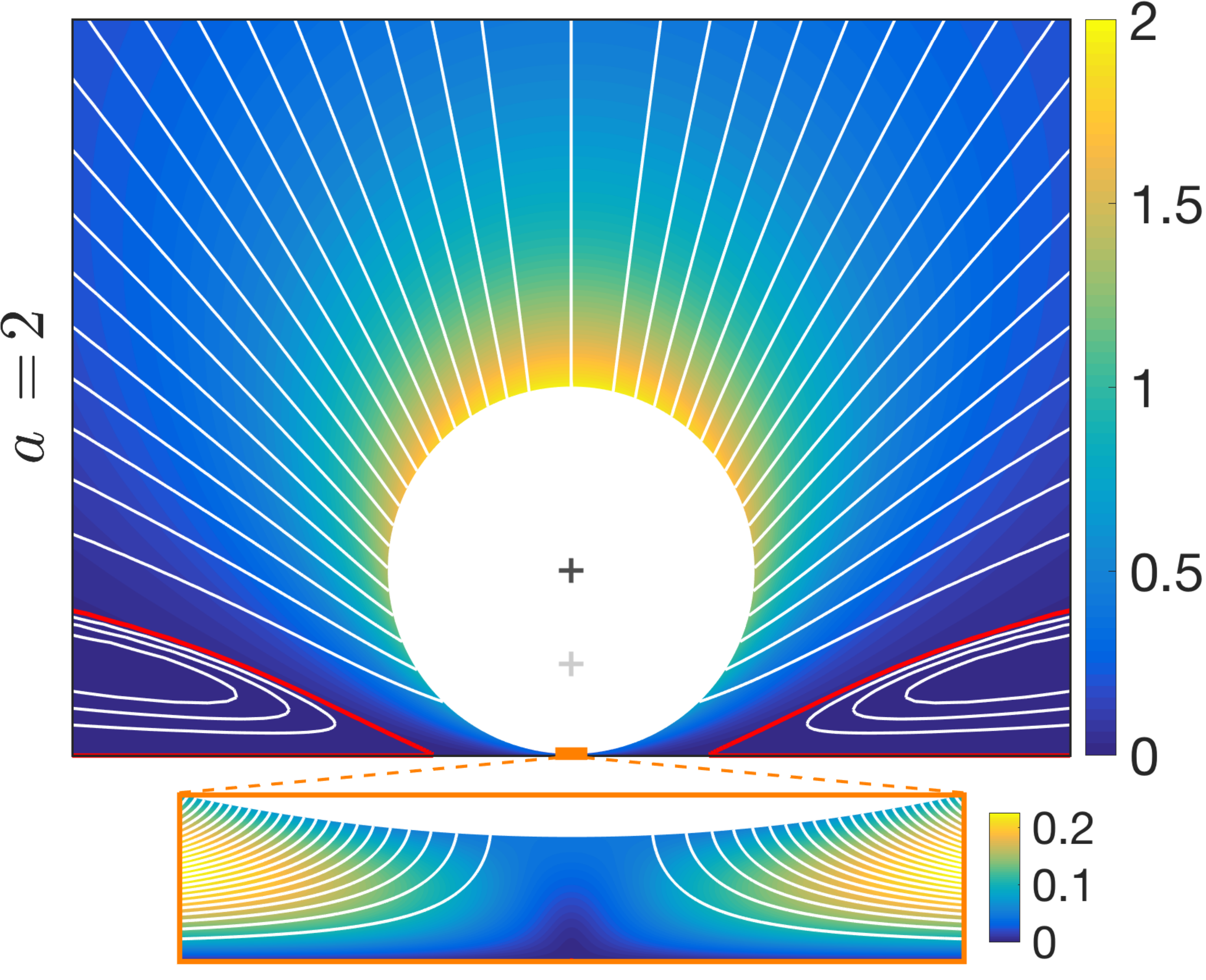} &
\includegraphics[width=.33\textwidth]{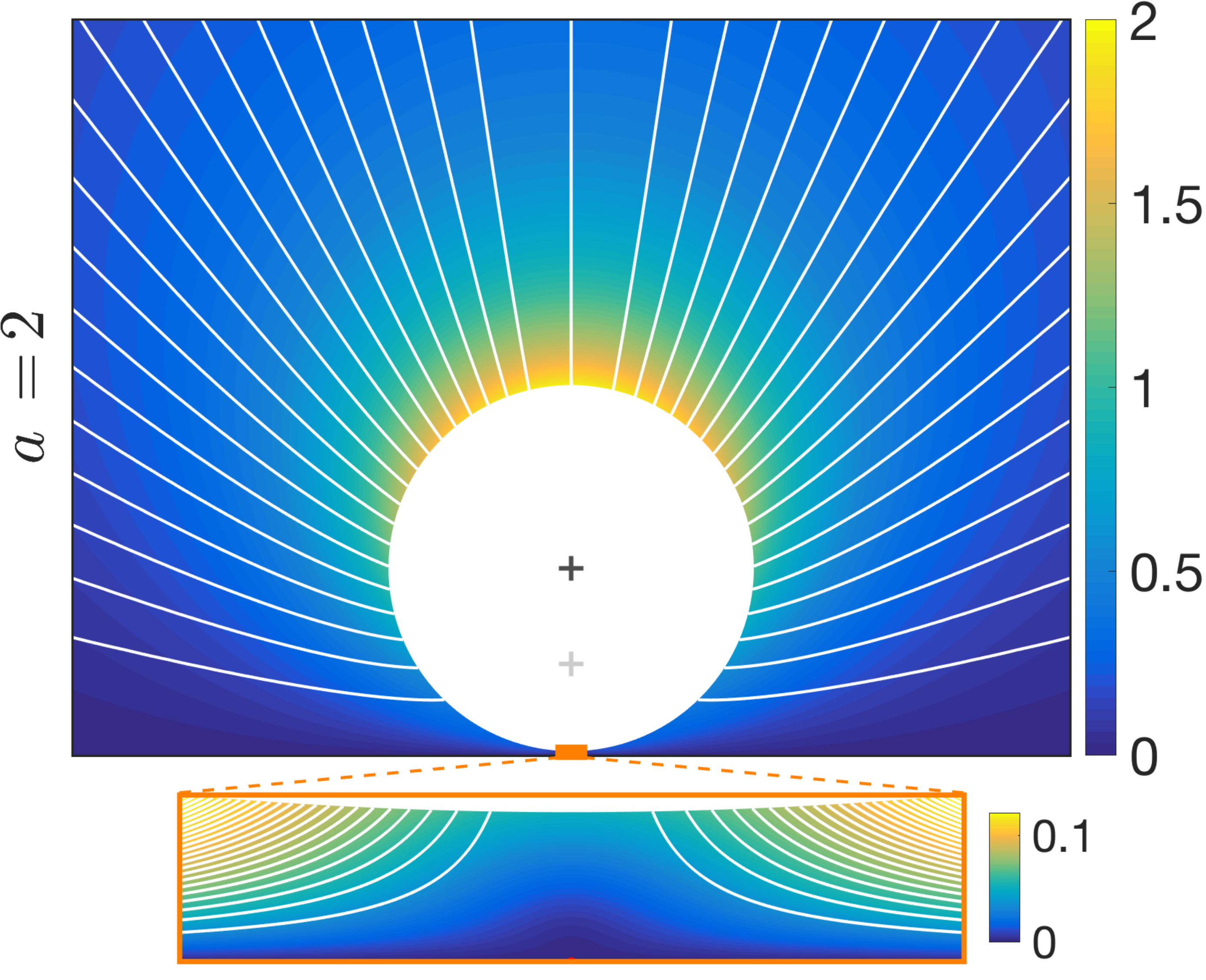} &
\includegraphics[width=.33\textwidth]{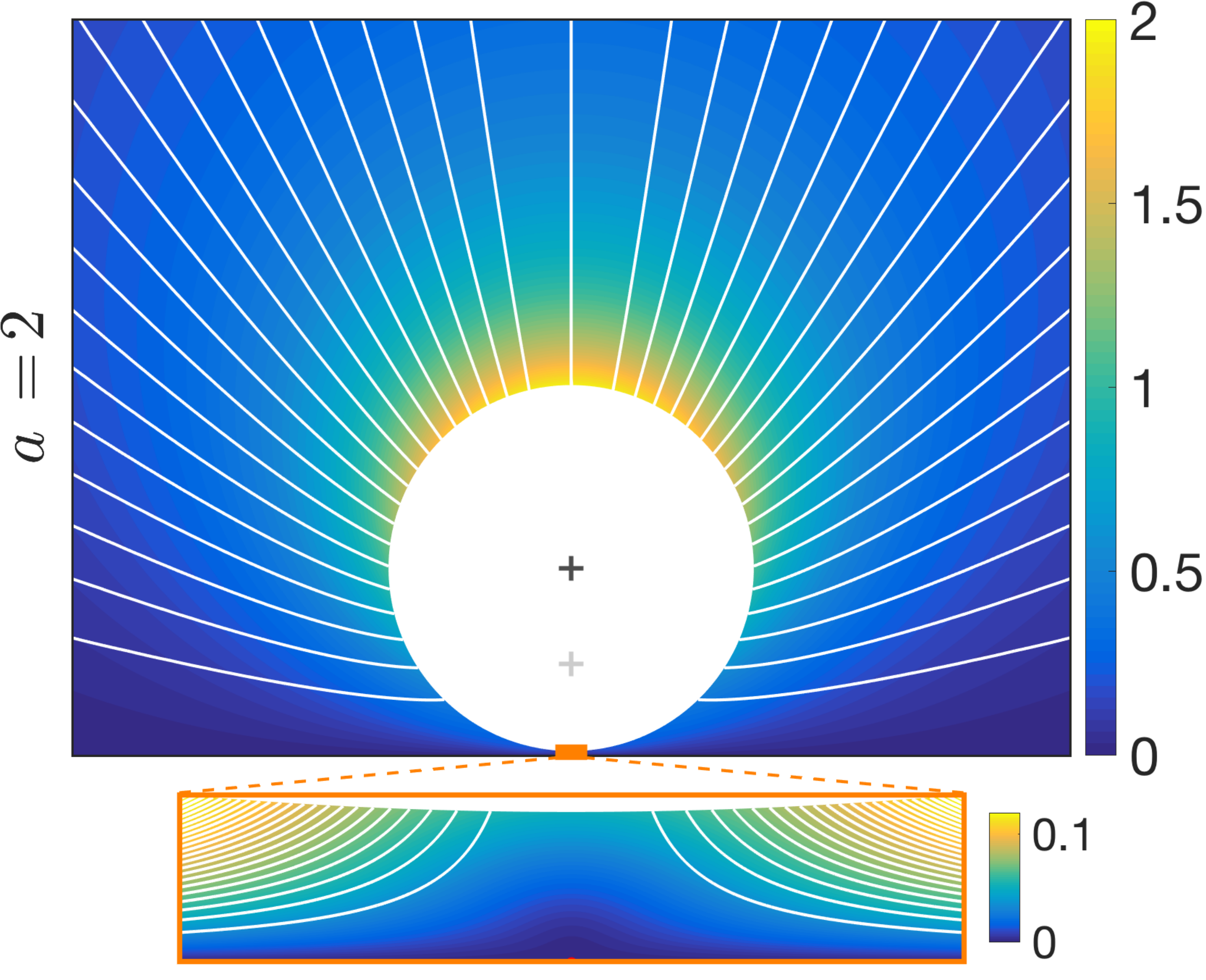}\\

&&\\
 &
\includegraphics[width=.33\textwidth]{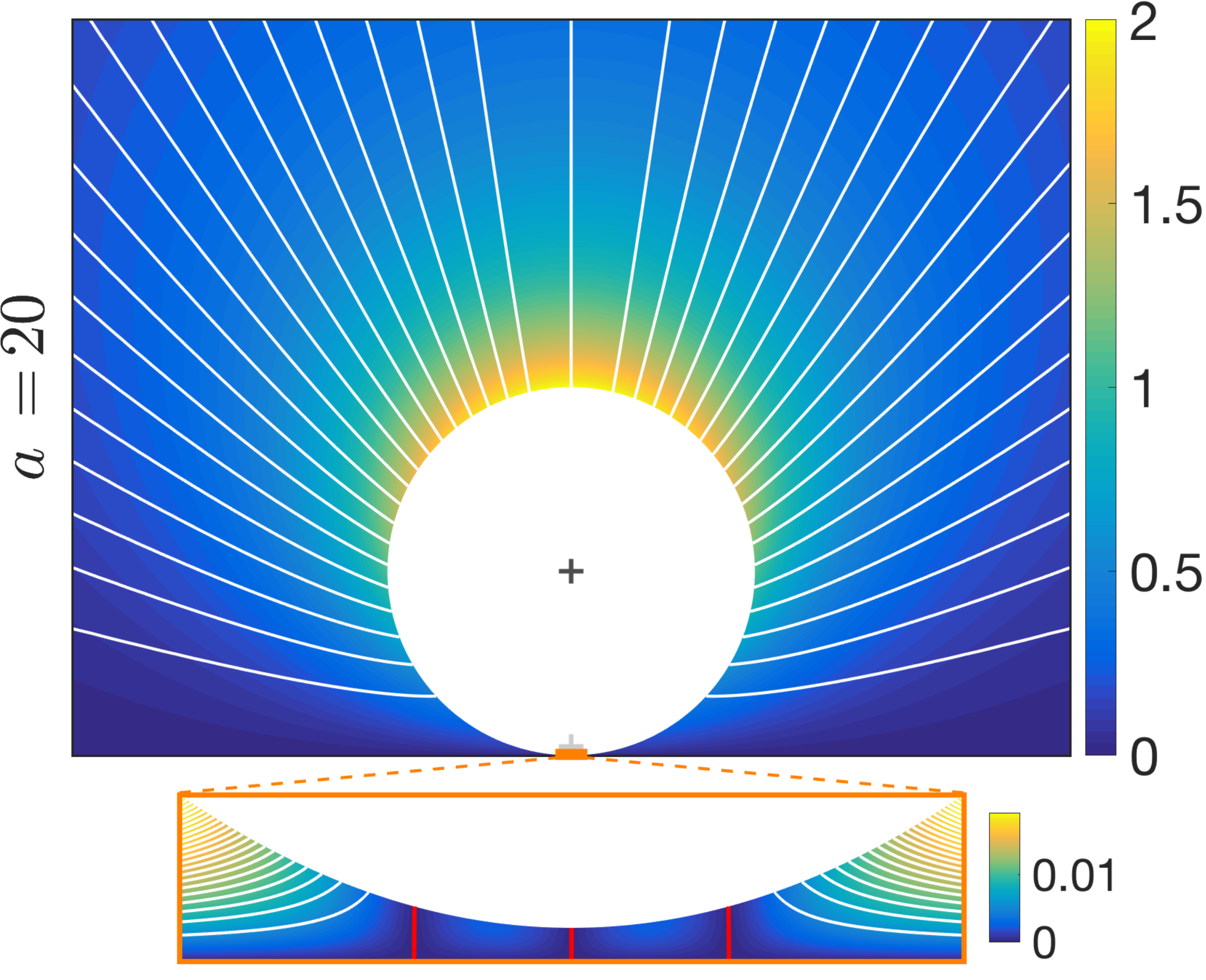} &
\includegraphics[width=.33\textwidth]{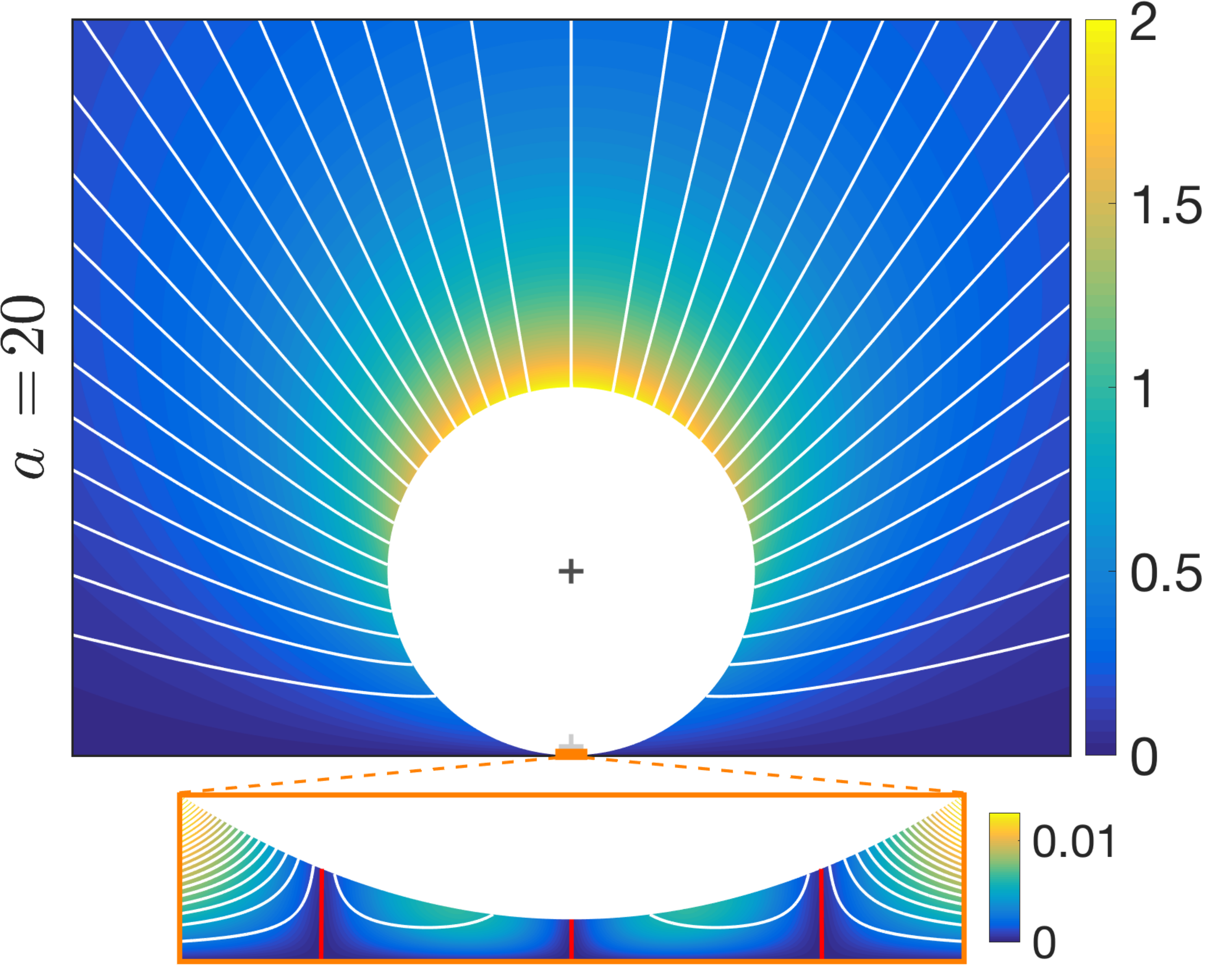}\\
&&\\
 &
\includegraphics[width=.33\textwidth]{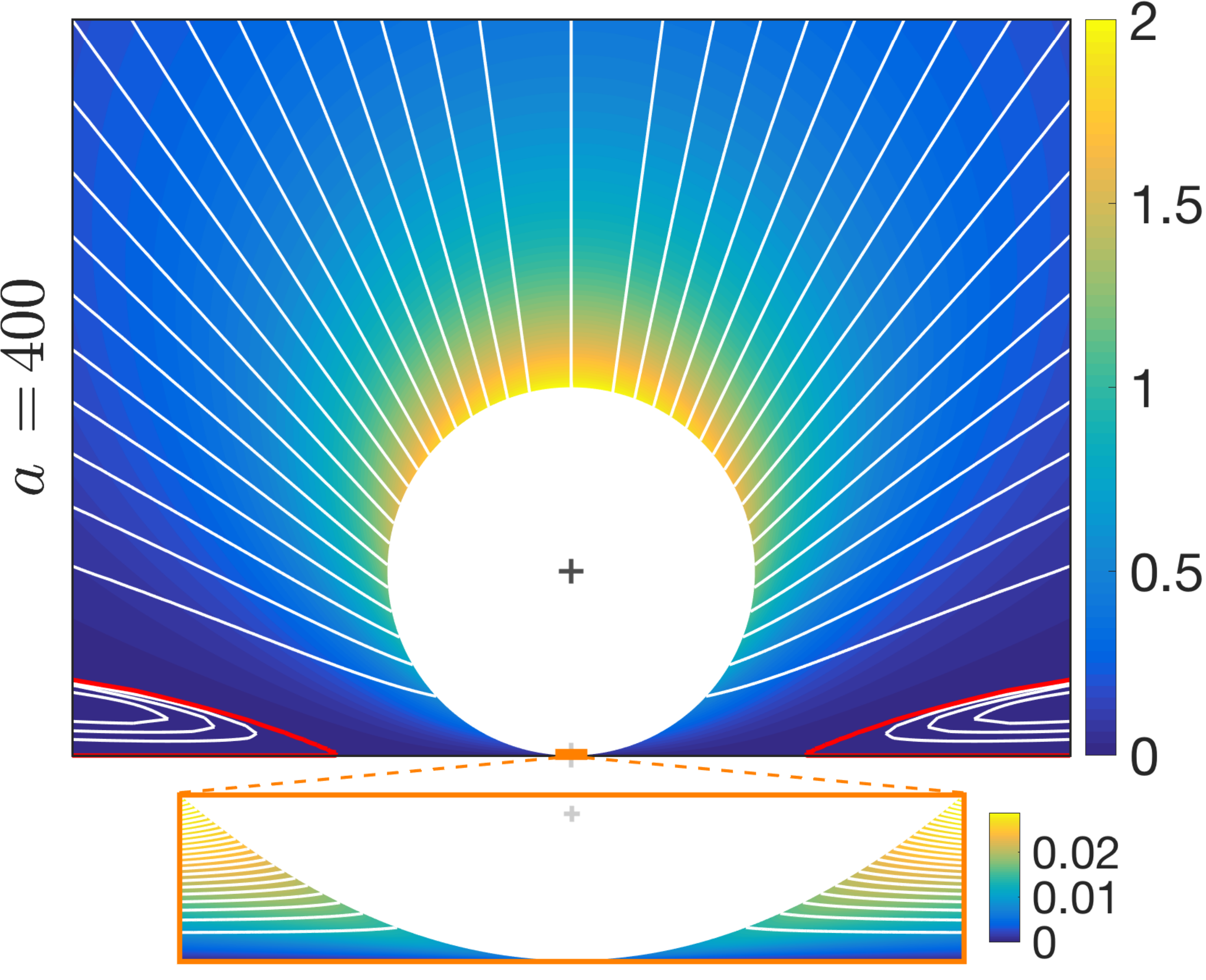} &
\includegraphics[width=.33\textwidth]{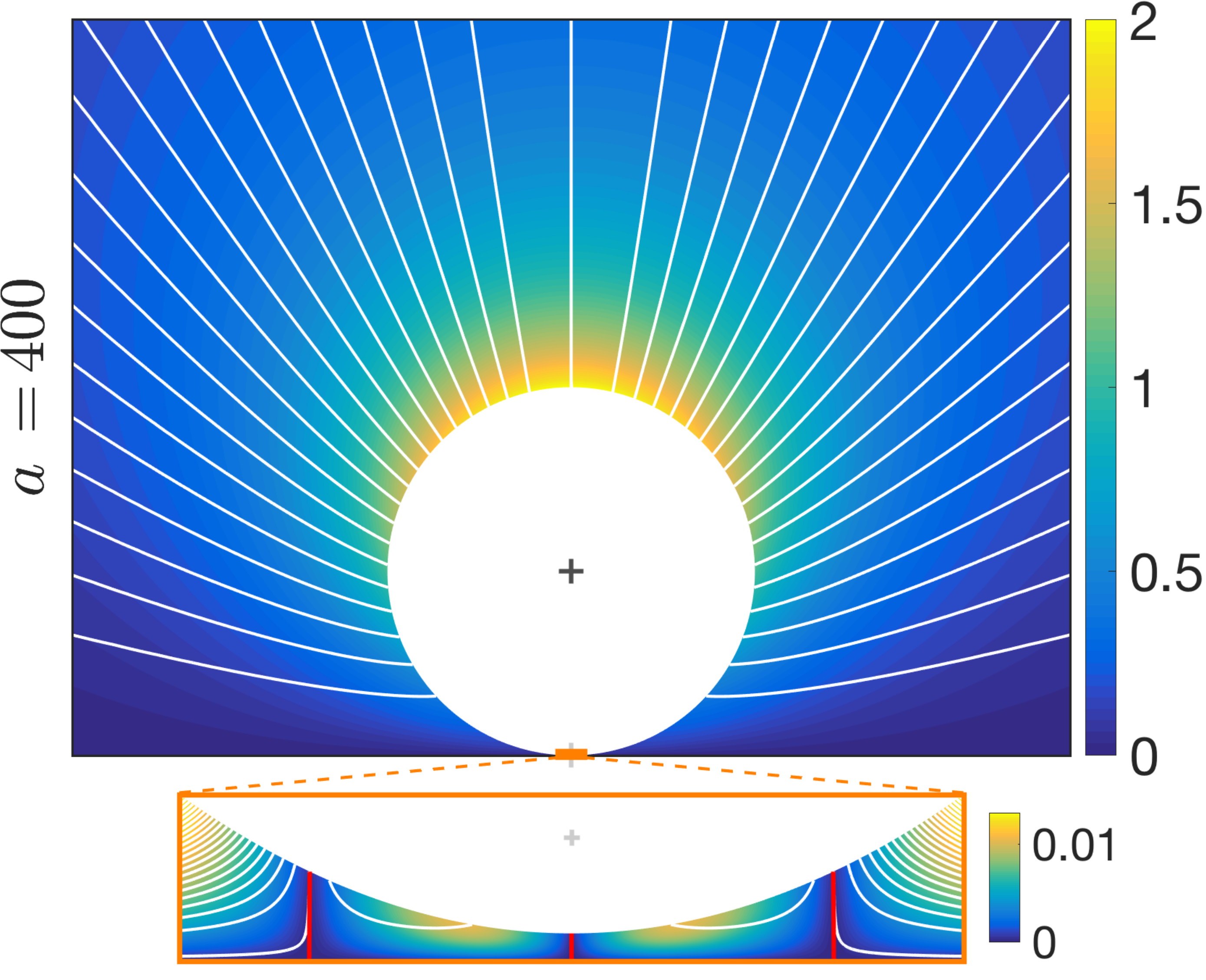}\\
\end{tabular}
\caption{Streamlines (white solid lines) and velocity magnitude (color) around the inflating bubble/sphere for (a) $\lambda/d_0=0$ (rigid shell), (b) $\lambda/d_0=50$ and (c) $\lambda/d_0=\infty$ (clean bubble), shown for $a/d_0=0.5$, $2$, $20$ and $400$ (from top to bottom). The current position of the bubble centre is shown as a black cross and the initial bubble position is shown for reference (light grey). Insets show the same information within the lubricating gap (the horizontal limits of the insets are $\pm 0.075a$, and vertical scale is stretched so as to show the entire gap width). In all panels, the streamline $\psi=0$ is highlighted in red indicating the boundary of recirculation regions.}\label{fig:flow}
\end{center}
\end{figure}

In figure~\ref{fig:flow}, we illustrate  the evolution of the flow field around the growing bubble (and within the lubricating film) throughout the growth process for three different cases: a rigid shell (no slip), a perfectly clean bubble (no-stress) and a finite-slip-length condition. The comparison of these three configurations provides  physical insight on the fluid dynamics processes involved both in the outer region around the bubble and within the thin fluid film in the lubrication limit. 

For all three boundary conditions, the initial growth of the bubble is qualitatively similar. Streamlines point radially out of the bubble and are tilted by the presence of the wall. Quantitative details of the flow field depend weakly on the value of $\lambda$ as only the non-isotropic flow is influenced by this property. This flow pattern corresponds to region (A) in figure~\ref{fig:growth_general}.

The three different cases differ however in their subsequent dynamics when the  thickness of the fluid layer becomes comparable or smaller than the radius of the  bubble. For the rigid shell, a lubricating film is established and forced by the displacement of the sphere's surface (remember that no slip is allowed here). As a consequence the fluid velocity is maximum in the centre of the gap and oriented outward as the film is draining. In the outer flow, a toroidal recirculation region is observed around the contact region and fluid is actually pumped (weakly) toward the inflating sphere just above the no-slip wall and at a finite distance  that scales with the radius of the sphere. This flow pattern is characteristic of region (C) of figure~\ref{fig:growth_general}.

The clean bubble limit presents a fundamentally different behaviour, once the lubricating film is established. Because the surface of the inflating bubble cannot sustain any tangential stress, it cannot force an outward flow within the lubricating film as it did in the case of a rigid shell. In the outer part of the film, the growth of the bubble indeed forces fluid out of the lubrication region but the flow is qualitatively different from the rigid shell limit as the fluid velocity reaches its maximum at the bubble's surface (instead of the centre of the lubrication film). This outward flow (associated with a downward motion of the bubble interface) is only present outside a cylindrical recirculation region (closed streamlines) whose dimensions scale linearly with the bubble size, and increases with $\lambda$. In that region, the net displacement of the bubble surface  is oriented upward. This flow pattern is characteristic of region (B) of figure~\ref{fig:growth_general}.

In the intermediate regimes where $\lambda$ is finite (but large enough for a rebound to occur), the three flow patterns associated with regimes (A), (B) and (C) are observed successively. While the fluid gap is small but $a$ is also sufficiently small (with respect to $\lambda$) the flow pattern observed for a bubble is found (recirculation within the lubricating film), while the final dynamics eventually returns to that of a rigid shell (recirculation region in the outer flow).

Note that figure~\ref{fig:flow} allows for a characterization of the three regimes identified in figure~\ref{fig:growth_general} in terms of flow field:
\begin{itemize}\renewcommand{\labelitemi}{$-$} 
\item[(A)]{Isotropic flow near the growing bubble and far from the wall, limited displacement of the bubble centre and bending of the streamlines by the confinement.}
\item[(B)]{Lubricating film with a recirculation region whose extent is proportional to the bubble radius, the ratio being an increasing function of $\lambda$. Outward flow in the rest of the lubricating film with maximum velocity at the surface of the bubble. Outward flow in the outer region.}
\item[(C)]{Lubricating film with an outward flow within the entire film, with maximum velocity at the centre. Toroidal recirculation in the outer region close to the wall and at a finite distance from the bubble (i.e.~its inner limit scales with the bubble radius).}
\end{itemize}
\vspace{.3cm}

In contrast with other classical problems involving rigid spheres and constant-volume bubbles in Stokes flow in close proximity to a rigid wall, the present \sm{results} identify the possibility for non-monotonic drainage of the lubricating film when the inflating bubble is increasing in size while remaining hydrodynamically force-free. \sm{Clean} bubbles \sm{(i.e. with perfect slip condition)} in fact experience a repulsion from the wall for small film thickness, in the sense that the absolute size of the film is increasing with $a$ (the relative gap width $d/a$ is however decreasing). For finite slip length $\lambda$, this repulsion regime is experienced provided that $\lambda$ is large compared to the bubble radius. For a sufficiently large  bubble, the boundary of the bubble behaves again as a rigid surface and the 
 thickness of the  lubrication film decreases exponentially.

\section{Flow forces on a translating and growing confined bubble}
\label{sec:forces}
In order to further understand the bubble dynamics  we  provide in this section a more in-depth analysis of the hydrodynamic force associated with the bubble growth. In particular, we analyse below (i) the influence of $\lambda$ on the hydrodynamic forces and (ii) the physical origin of the hydrodynamic repulsion from the wall associated with bubble growth. These results will be further  confirmed by considering the leading-order hydrodynamic contributions in the thin film  using lubrication analysis in Section~\ref{sec:lub}.

The kinematics of the bubble motion is fully described by the rate of change in bubble radius, $\dot{a}$, and in minimum bubble-to-wall distance, $\dot{d}$. Since the Stokes equations are linear and time-independent, the contribution to the hydrodynamic forces of each of these components (inflation and translation)  can be obtained separately and superimposed at each instant, Eq.~\eqref{eq:force}, namely we may write
\begin{equation}
F_I(\varepsilon,\tilde\lambda)=C_I(\varepsilon,\tilde\lambda)a\dot{a},\qquad F_T(\varepsilon,\tilde\lambda)=C_T(\varepsilon,\tilde\lambda)a\dot{d}.
\end{equation}

The pure translation situation leading to the definition of $C_T$ (i.e.~$\dot{a}=0$) is a classical problem, both in unbounded flows (Stokes drag) and under confinement \citep{kimbook,leal}. It is well known that the force $F_T$ will oppose the motion of the body, hence $C_T<0$ regardless of the value of $\varepsilon$ and $\tilde\lambda$. The force coefficient diverges as $\sim\varepsilon^{-1}$ for small gap width, regardless of the value of $\lambda$. This divergence is physically associated with the change in film thickness near the axis where fluid needs to be brought within the lubricating gap to achieve an $O(1)$ vertical velocity at the axis, which requires large longitudinal pressure gradients within the gap. 

Any non-monotonic behaviour arises therefore from the variations  in direction of the fluid force $C_I$ associated with the inflation of the bubble for fixed gap width $d$. In the small-gap limit, inflation of the bubble drives a flow within the lubrication film as the bubble radius increases, but in a much smaller extent as the vertical velocities are now $O(\varepsilon)$. Hence, the leading-order scaling of $C_I$ is expected to be weaker than $O(\varepsilon^{-1})$ for $\varepsilon\ll 1$. 

In the following, we first present and analyse the full solution for the two components of the hydrodynamic forces and associated flow field as obtained from the complete solution (Section~\ref{sec:model}) before turning to the lubrication analysis within the fluid gap for small $\varepsilon$ in Section~\ref{sec:lub}.

\subsection{Flow forces for arbitrary gap width}
The framework obtained in Section~\ref{sec:model} provides a semi-analytical definition of the force coefficients $C_T$ and $C_I$, respectively associated with the translation of the bubble (for fixed radius $a$) and its growth/collapse (for fixed gap width $d$), for arbitrary slip length~$\lambda$. 

\renewcommand{\labelitemi}{$-$}

\begin{figure}
\begin{center}
\begin{tabular}{cc}
\includegraphics[width=.49\textwidth]{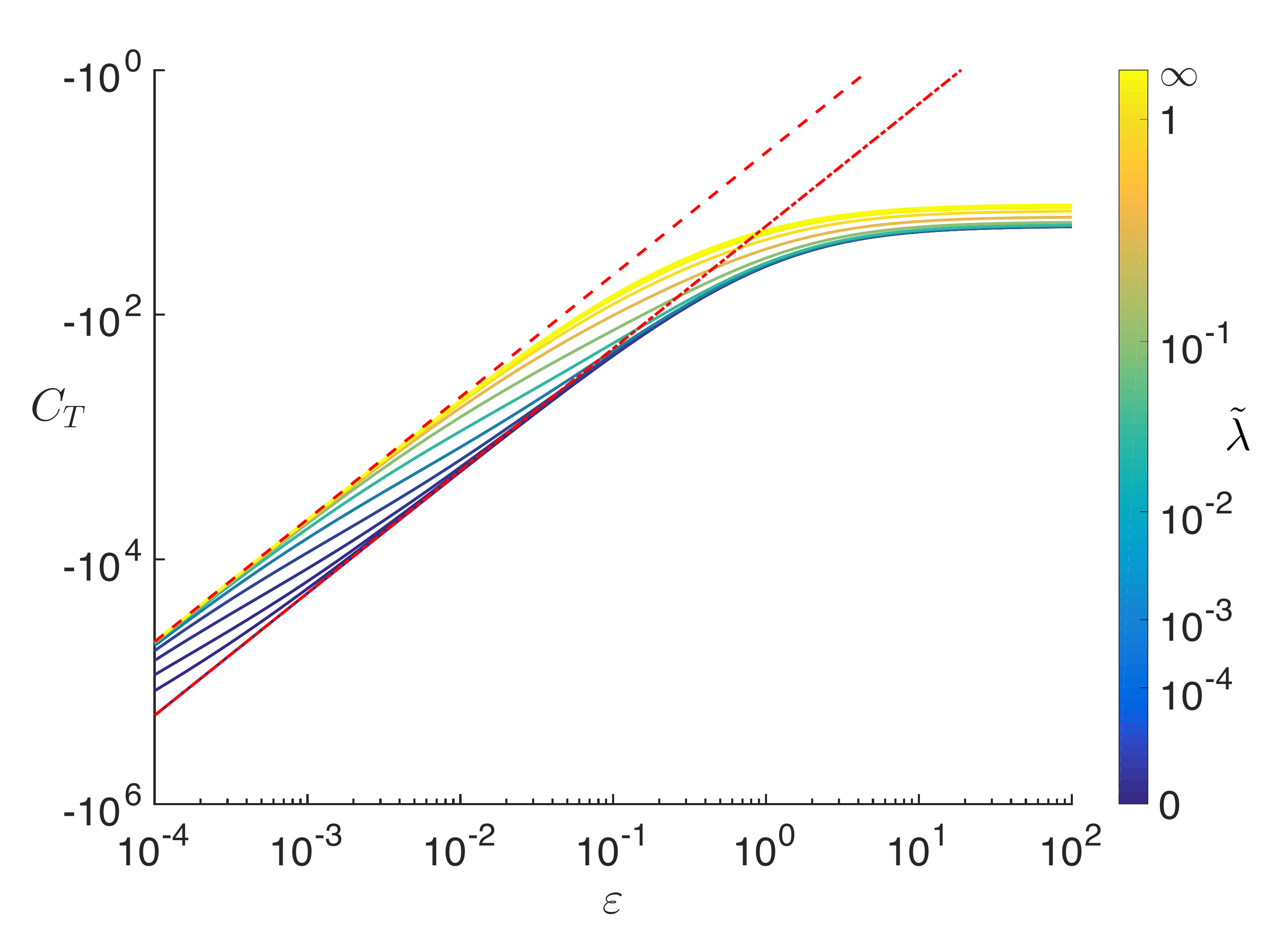} &
\includegraphics[width=.49\textwidth]{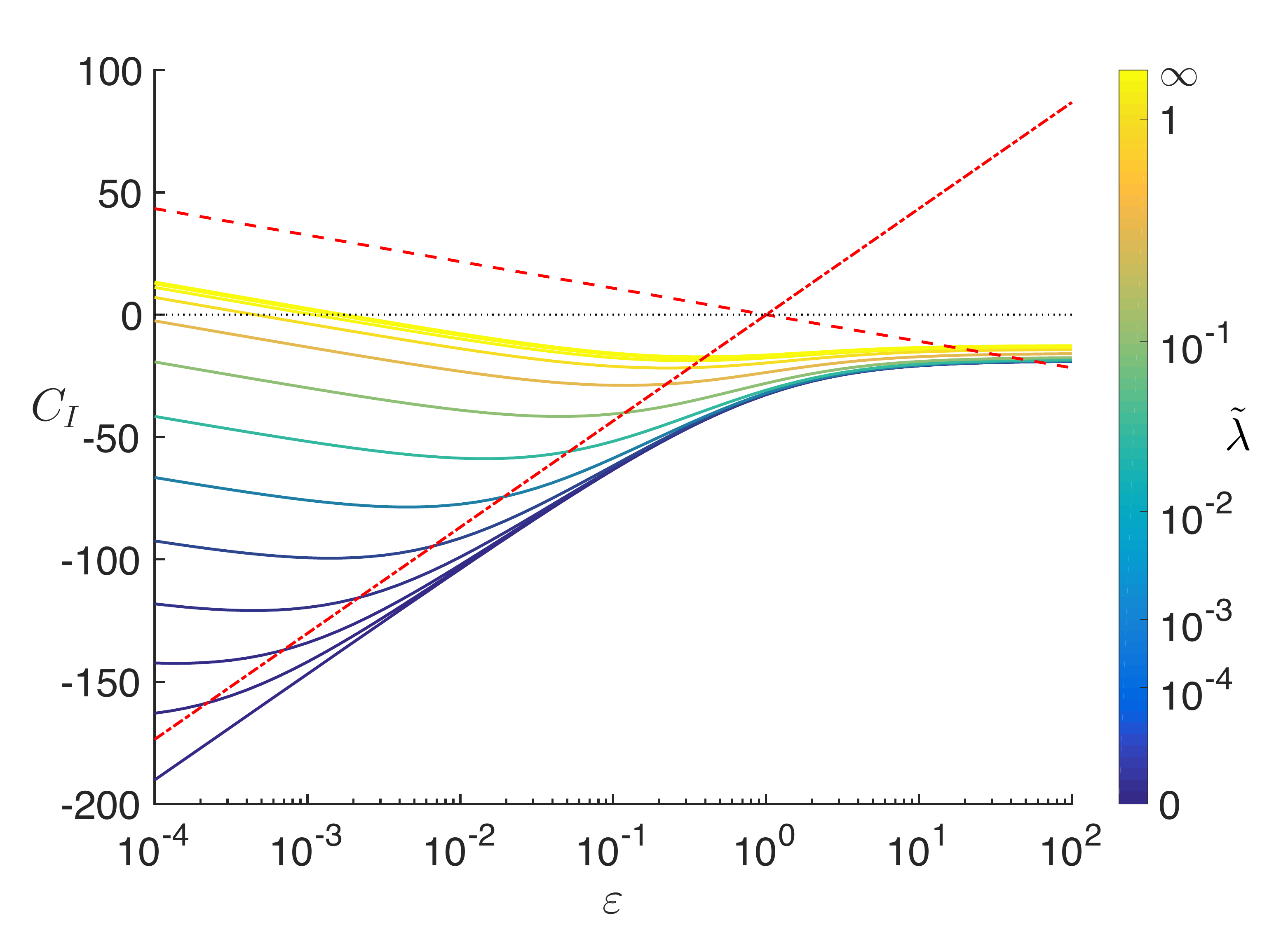}
\end{tabular}
\caption{Dependence of the force coefficients associated with  upward translation of the bubble (fixed radius $a$, left)  and  growth of the bubble (fixed gap width $d$, right) on  the relative gap width, $\varepsilon=d/a$, and reduced slip-length, $\tilde\lambda=\lambda/a$. The leading-order asymptotic approximations obtained in Section~\ref{sec:lub} for a perfectly clean bubble ($\tilde\lambda=\infty$) and rigid spherical shell ($\tilde\lambda=0$) are shown in red dashed and dash-dotted lines respectively.}\label{fig:force_general}
\end{center}
\end{figure}

Their evolution with the relative gap width $\varepsilon=d/a$ and reduced slip length $\tilde\lambda=\lambda/a$ is reported on figure~\ref{fig:force_general}. Note that by convention  $C_T>0$ (resp.~$C_T<0$) corresponds to a hydrodynamic force pointing away from (resp.~toward) the wall for a bubble translating away from the wall (i.e.~$\dot{d}>0$ or upward), and
 $C_I>0$ (resp.~$C_I<0$) corresponds to a hydrodynamic force pointing away from (resp.~toward) the wall for a growing bubble ($\dot{a}>0$). As for $C_T$, $C_I<0$ corresponds to the intuitive resistance behaviour of Stokes flow to the general upward motion of the bubble (for $\dot{d}=0$, the upper half and most of the rest of the bubble surface is moving upward, pushing fluid in that direction).

Inspecting the results in figure~\ref{fig:force_general},  we see that the translation of the bubble always results in a resistive hydrodynamic force ($C_T<0$) that diverges when the bubble is close to the wall, as expected. The scaling in $\sim 1/\varepsilon$ is consistent with classical results on the dynamics of thin viscous fluid films~\citep{leal,kimbook}. For the limiting cases of a perfectly-clean bubble ($\tilde\lambda=\infty$) and rigid spherical shell ($\tilde\lambda=0$), the results are found in excellent quantitative agreement with the leading-order asymptotic expansion provided by the lubrication analysis presented in Section~\ref{sec:lub} (see also Ref.~\citep{leal}). In particular,  the magnitude of the drag on a clean bubble is  a quarter of that on a spherical shell. 

For $0<\tilde\lambda<\infty$ (i.e.~intermediate slip lengths), the behaviour of the force coefficient transitions from one limit to the other for $\varepsilon\sim\tilde\lambda$. This can be understood as follows: the hydrodynamic force on the bubble (or at least its correction to its unbounded value) is dominated by the effect of the wall, and the physical behaviour of the boundary is conditioned by the relative magnitude of the slip length $\lambda$ and the distance to the wall $d$. For $d\ll \lambda$ (i.e.~$\varepsilon\ll \tilde\lambda$), the surface of the bubble appears effectively clean and the problem is identical to that of the perfect-slip limit $\lambda=\infty$, while for $d\gg \lambda$ (i.e.~$\varepsilon\gg \tilde\lambda$), the surface of the bubble is effectively rigid and the rigid shell limit is recovered.

The force coefficient associated with bubble growth ($C_I$) displays more complex variations. Its general scaling in $\log\varepsilon$ diverges more slowly than $C_T$ for small gap, as expected since the minimum gap width is held fixed. Here, the no-slip ($\tilde\lambda=0$) and perfect-slip ($\tilde\lambda=\infty$) limits display fundamentally different behaviours. The no-slip case results in a \sm{negative} hydrodynamic force \sm{(with $\dot{a}>0$)}, which increases in magnitude as the gap width $\varepsilon$ is reduced. This is the result of the superposition of two intuitive effects, namely the \sm{$O(1)$ negative hydrodynamic force resisting the upward motion of the upper boundary of the sphere (the gap width being held fixed, $\dot{a}>0$ imposes a general upward motion of most of the bubble surface)} and the resistance to the lubrication film drainage associated with the increase in bubble radius (albeit much smaller than what is observed for translation). The latter accounts for the logarithmic divergence of $C_I$ for small $\varepsilon$. 

In the opposite clean-bubble limit, the hydrodynamic force is shown to increase with decreasing gap width, eventually becoming positive for $\varepsilon<\varepsilon_c^{\tilde\lambda=\infty}\approx 1.8\times10^{-3}$. For small gaps, a growing bubble experiences therefore a hydrodynamic force that pushes it away from the wall. This can  be understood qualitatively as the result of the superposition of two competing \sm{effects}: an $O(1)$ hydrodynamic resistance associated with the upward motion of the outer boundary of the bubble (as above), and a singular (positive) force associated with the film dynamics. 

Intermediate cases $0<\tilde\lambda<\infty$ show  again a transition between the two regimes around $\varepsilon\approx\tilde\lambda$: in that configuration, the resistance to bubble growth ($C_I<0$) exhibits a  maximum in magnitude for a finite distance. 

\subsection{Flow field between a growing bubble and a wall for a fixed fluid gap}
\label{sec:inflating}
\begin{figure}
\begin{center}
\begin{tabular}{cc}
Rigid shell ($\lambda=0$) & Clean bubble ($\lambda=\infty$)\\
\includegraphics[width=.48\textwidth]{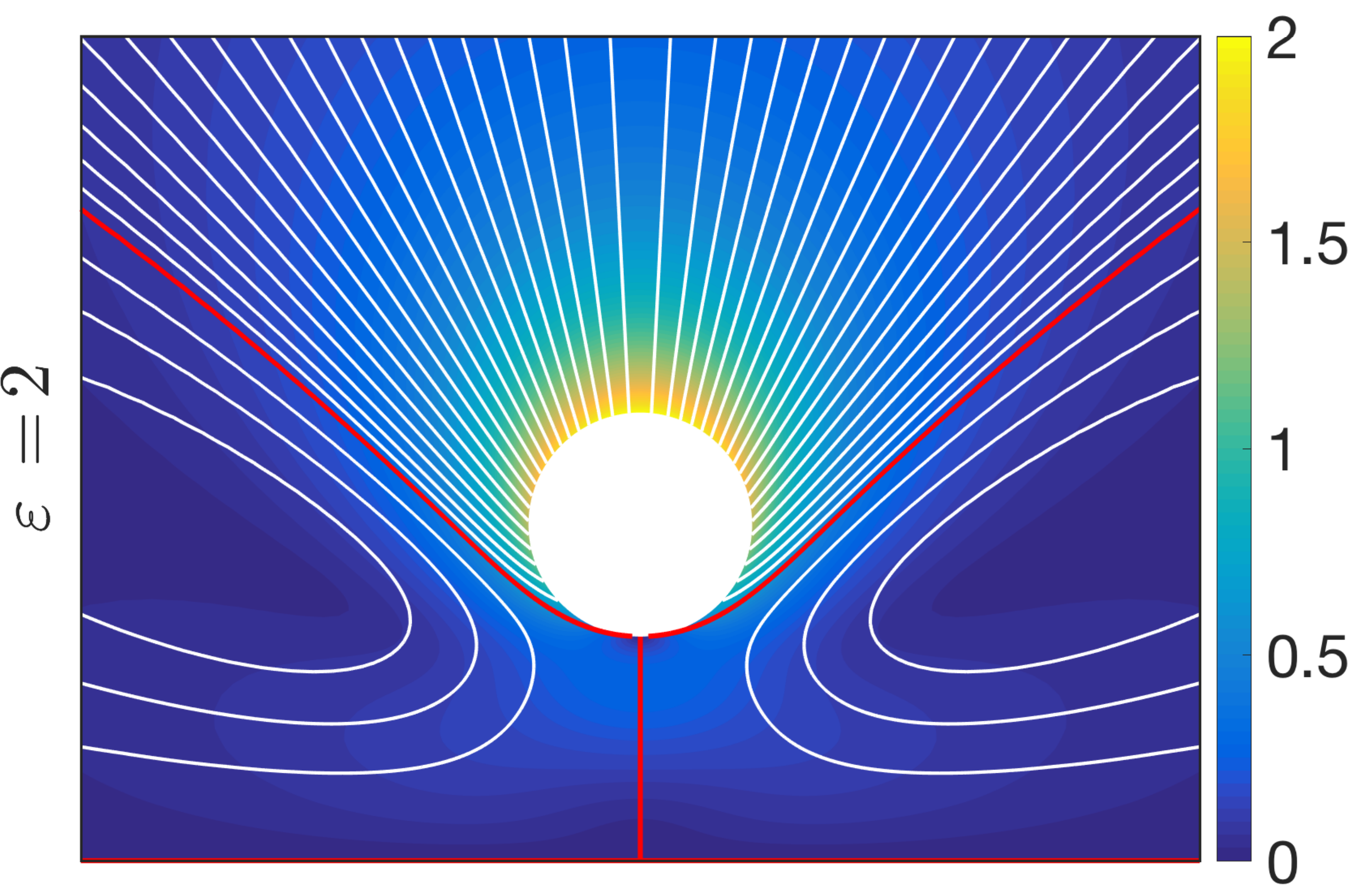} &
\includegraphics[width=.48\textwidth]{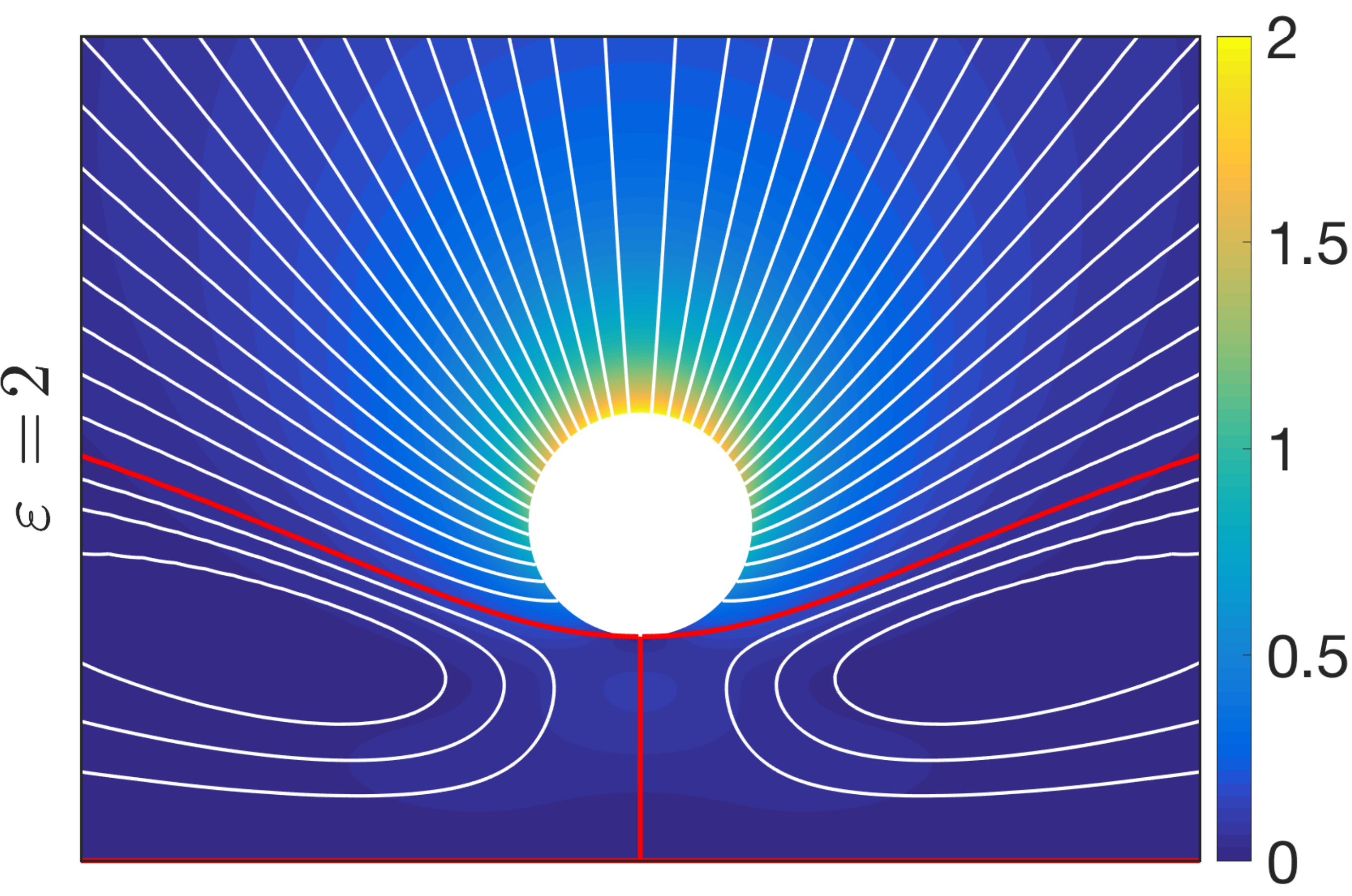}\\
&\\
\includegraphics[width=.48\textwidth]{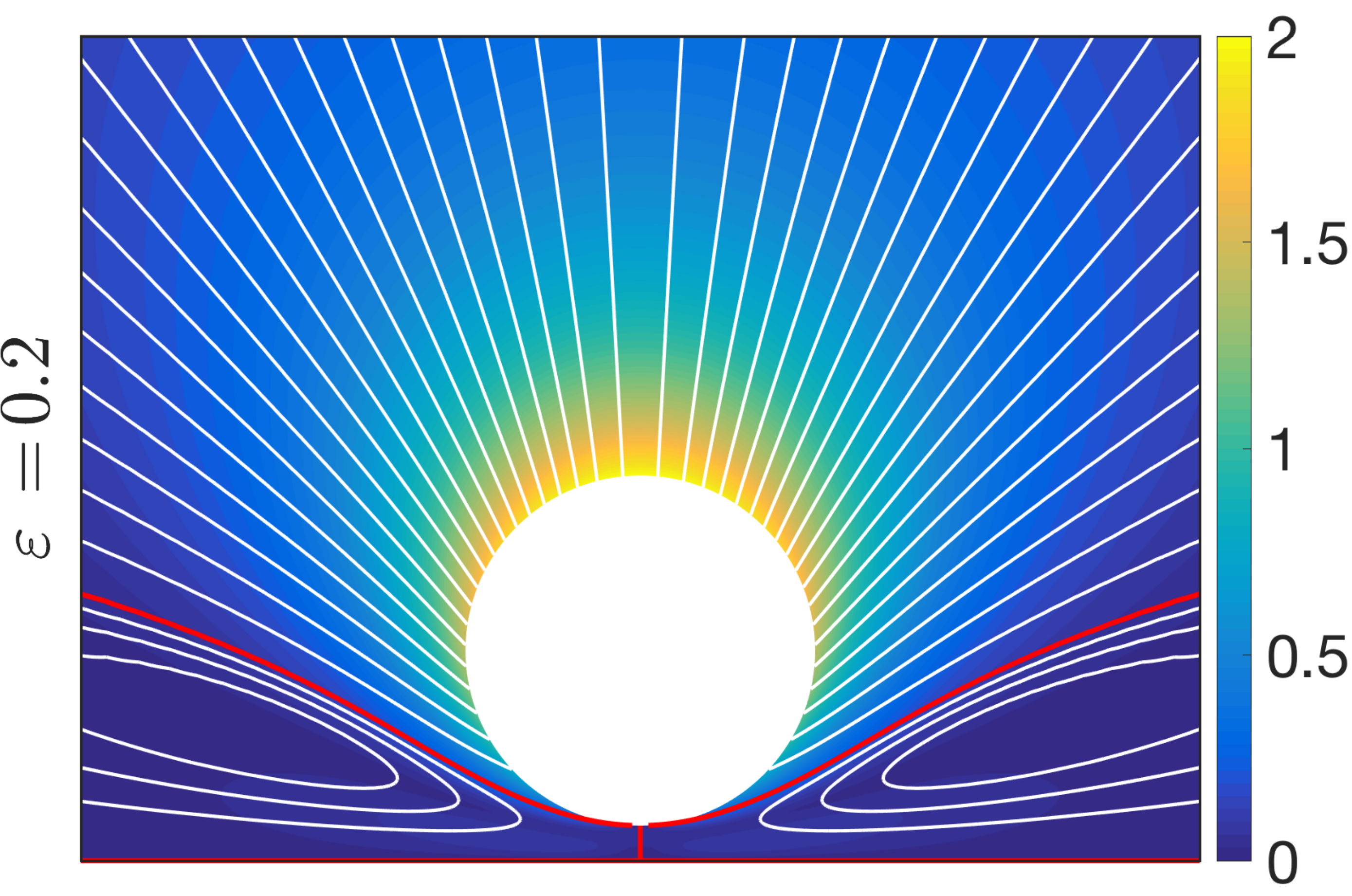} &
\includegraphics[width=.48\textwidth]{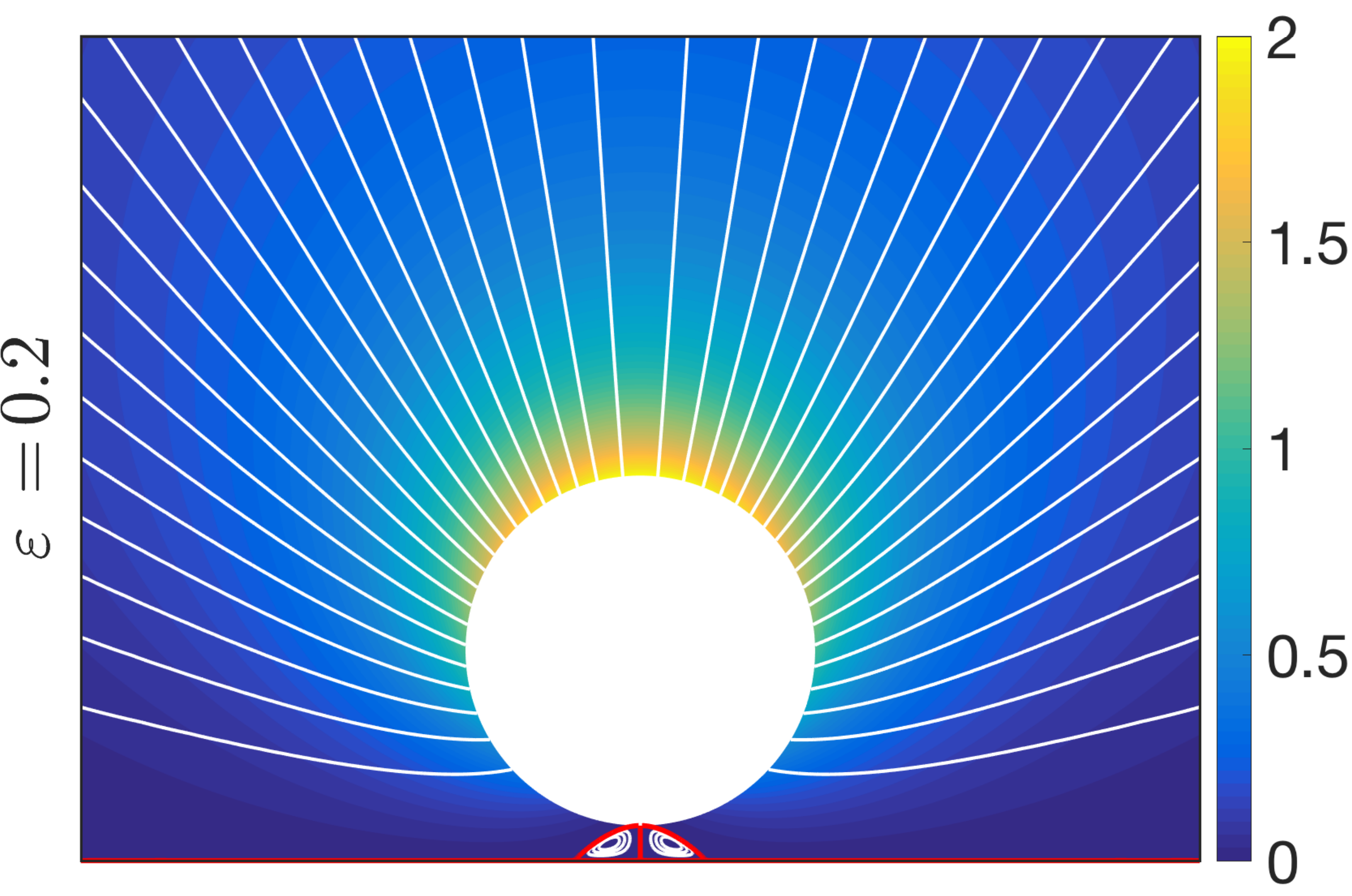}\\
&\\
\includegraphics[width=.48\textwidth]{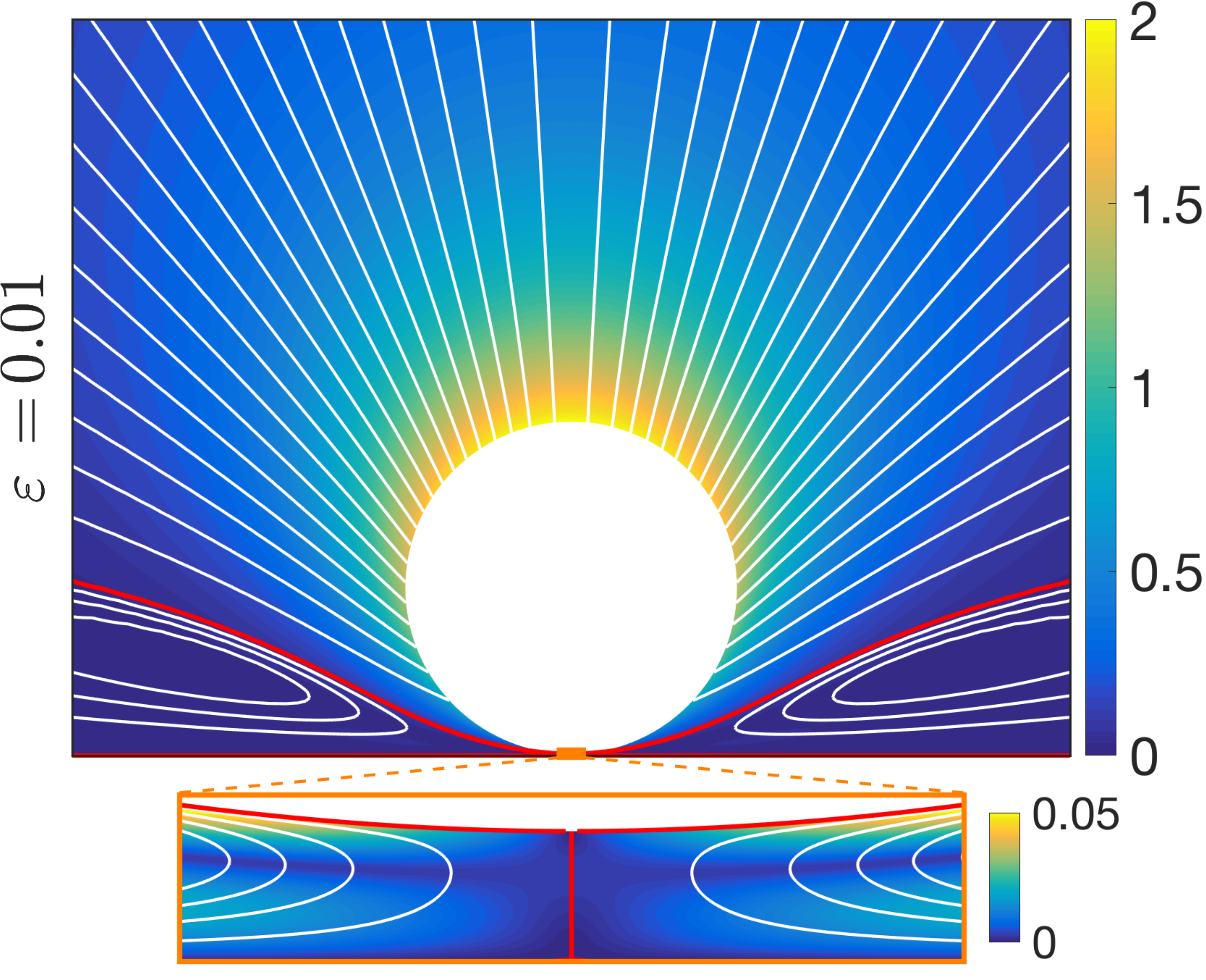} &
\includegraphics[width=.48\textwidth]{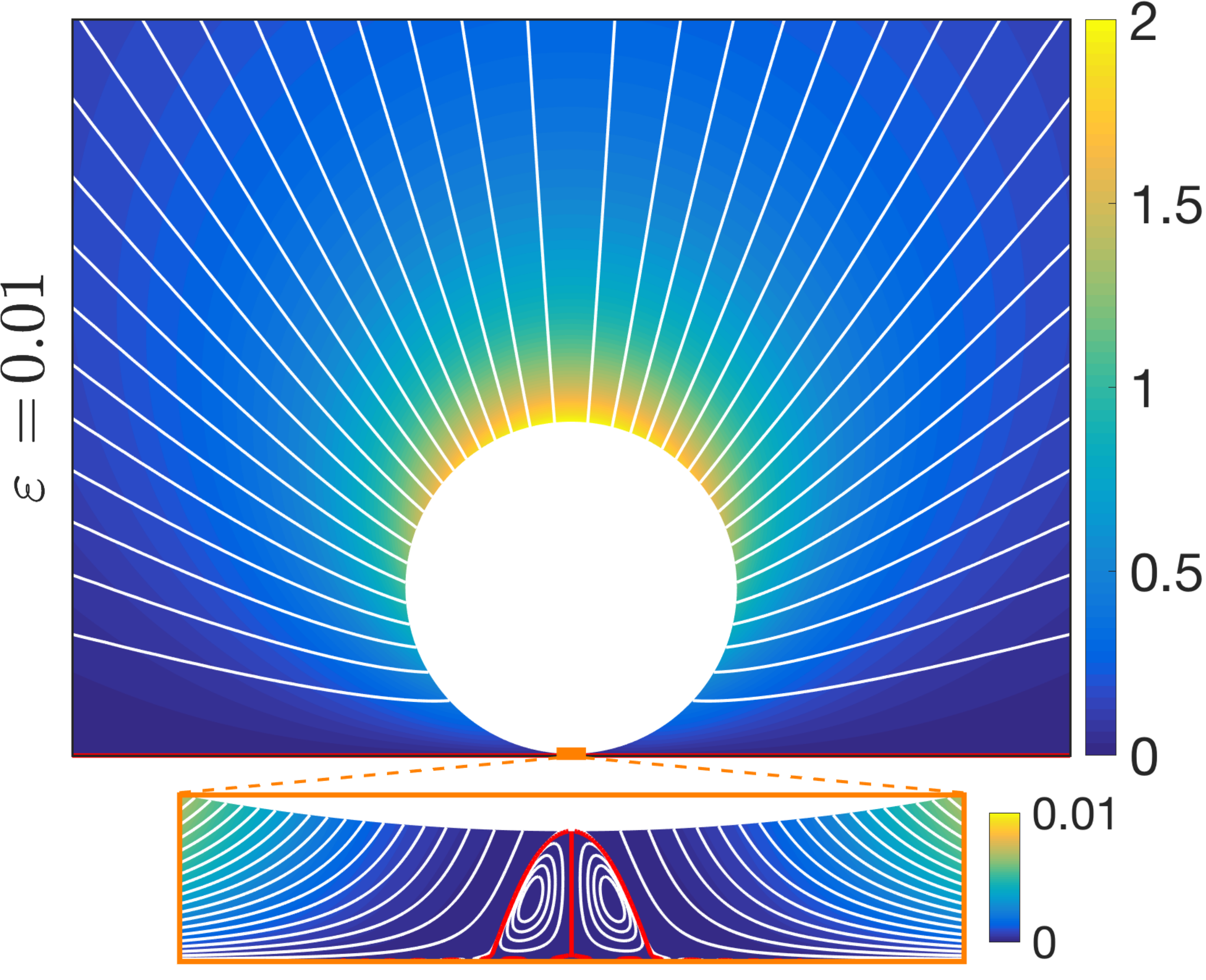}
\end{tabular}
\caption{Streamlines (white lines) and velocity magnitude (color) for the flow generated around an inflating rigid shell ($\lambda=0$, left) and a clean bubble ($\lambda=\infty$, right) for fixed gap width (top) $\varepsilon=2$, (centre) $\varepsilon=0.2$ and (bottom) $\varepsilon=10^{-2}$. For $\varepsilon=0.01$, the inset shows the same information within the lubricating gap (the horizontal limits of the inset are $\pm 0.075a$, and the vertical scale is stretched so as to display the entire gap width). In all panels, $a=1$, $\dot{a}=1$ and $\dot d=0$, and dividing streamlines between recirculating and expanding regions are shown in red.}
\label{fig:flow_macro}
\end{center}
\end{figure}

The  surprising result  of the previous analysis is that the force experienced by a growing bubble near a wall  can, {when the fluid gap width is held fixed},  \sm{reverse} sign depending on the wall proximity and the value of the slip length. As a result,
 a force-free growing clean bubble with a thin lubrication gap is repelled by the confining wall, in contrast with what happens for a rigid shell or for weaker confinement.  This reversal stems from a fundamentally-different behaviour of the flow field  for an effectively rigid interface ($\lambda\ll d$) and for an effectively stress-free one ($\lambda\gg d$), which we illustrate in figure~\ref{fig:flow_macro} where the streamlines and velocity magnitude are shown for the two limiting cases of surfaces with zero and infinite slip length and varying gap width.

Let us first consider as a reference the case of an inflating bubble or sphere far from the wall. Holding its bottom boundary fixed leads in both cases to the formation of a recirculation torus near the wall and around the axis of symmetry of the problem, pumping fluid inward along the wall. Outside this recirculation region, the flow is oriented outward forced by the radial displacement of the bubble's boundary.  Far from the wall, the topology of the recirculating regions is roughly similar for both $\lambda=0$ and $\lambda=\infty$, with the main difference being only a weaker flow velocity behind a clean surface which allows for fluid slip and is therefore not able to drive a tangential fluid motion.

The two limits $\lambda=0$ and $\lambda=\infty$ differ however fundamentally in the behaviour of this recirculation zone when the distance to the wall is reduced. In the case of a no-slip boundary ($\lambda=0$), a recirculation torus whose size scales with the radius of the bubble is maintained around the bubble and within the thin lubricating film in the limit $\varepsilon\rightarrow 0$. This is only possible because the surface of the sphere  can sustain a net shear stress and continues driving the fluid tangentially to its surface and out of the gap. Note that this recirculation region leads to a reversal of the flow field within the lubrication gap, which imposes the pressure within the lubricating film to be smaller than the outer pressure (see Section~\ref{sec:lub}). As a result, the effect of the lubricating film is to pull the inflating bubble downward (i.e.~$C_I<0$).

In contrast, for a no-stress boundary ($\lambda=\infty$), the recirculation torus still exists but it only scales with the lubrication gap width, since the surface of the  bubble can only exert normal stresses on the fluid and cannot counter act the effect of the no-slip wall. In the limit $\varepsilon\rightarrow 0$, it is therefore confined to the immediate vicinity of the axis and its impact on the total force on the bubble all but disappears. The flow within the lubricating film is now unidirectional and the lubricating pressure is higher than that in the outer flow leading to a hydrodynamic repulsion  ($C_I>0$).

\section{Asymptotic analysis of the flow and forces in the lubrication limit}
\label{sec:lub}
The   results above hint at the fundamental role played by the structure of the flow within the lubricating gap. We now confirm quantitatively this physical picture  by analysing the asymptotic behaviour of the flow within the gap  using lubrication theory in the asymptotic limit $\varepsilon\ll 1$. 

\subsection{Lubrication equations}

In the limit of small gap width, $\varepsilon\ll 1$, classical lubrication theory can  be used to solve for the flow field within the thin layer of fluid separating the inflating bubble from the wall. This thin film is described in polar coordinates by the location of the bubble's surface $z=h(r,t)$, given exactly instantaneously by
\begin{equation}
h(r,t)= d(t)+a(t)-\sqrt{a(t)^2-r^2}.
\end{equation} 
Without any approximation on the thickness of the film, the impermeability condition at the bubble surface is given by
\begin{equation}
u_z(r,h)=\pard{h}{t}+u_r(r,h)\pard{h}{r}\cdot\label{eq:imp_lub}
\end{equation}
Similarly, using the definition of the tangential and normal unit vectors $\mathbf{t}$ and $\nb$,
\begin{equation}
\mathbf{t}=\frac{\eb_r+h'\eb_z}{\sqrt{1+h'^2}},\qquad \nb=\frac{h'\eb_r-\eb_z}{\sqrt{1+h'^2}},
\end{equation}
and that of the reference surface velocity $\ub^\textrm{ref}=(\dot{a}+\dot{d})\eb_z+\dot{a}\nb$, Eq.~\eqref{eq:dynbc} is given   in the general case by
\begin{align}
h'\sqrt{1+h'^2}\left(\dot{d}+\dot{a}\right)=&(u_r+h'u_z)\sqrt{1+h'^2}+\lambda\left[(1-h'^2)\left(\pard{u_z}{r}+\pard{u_r}{z}\right)+2h'\left(\pard{u_z}{z}-\pard{u_r}{r}\right)\right].\label{eq:dynbcgen}
\end{align}
In the lubrication assumption, the typical film thickness $h$ is much smaller than the scale of its variations in the radial direction and so is its slope.  In that limit,
\begin{equation}
h(r,t)=d+\frac{r^2}{2a(t)}+O\left(\varepsilon^{3/2}\right),\label{eq:h_lub}
\end{equation}
and the lubrication framework is classically only valid within a $O(\sqrt{ad})$ region located near the axis of symmetry~\citep{leal}. Therefore, in the lubrication limit $h\sim d$ and $r\sim d\varepsilon^{-1/2}$.

Stokes' equations are linear and time-independent. The lubrication problem considered here can therefore be   solved as the superposition of two instantaneous and independent problems, respectively associated with the bubble translation (forced by $\dot{d}$ with $a$ fixed) and growth (forced by $\dot{a}$ with $d$ fixed). One could treat these two problems completely independently, and they are only presented jointly here for brevity, as the derivation follows the same structure.

Denoting by $V$ the typical leading-order vertical velocity scale for each problem, and using the fact that the  fluid incompressibility  imposes $u_r\sim\varepsilon^{-1/2}u_z$ at leading order, different scalings are obtained for the vertical velocity in each problem. Specifically,  $V\sim \dot{d}$ for the translation problem and $V\sim\varepsilon \dot{a}$ for the inflation problem. Keeping only the dominant terms for each problem, the leading-order contribution to the dynamic boundary condition~\eqref{eq:dynbcgen}   becomes
\begin{equation}
\lambda \pard{u_r}{z}+u_r=h'\dot{a}.\label{eq:dynbclub}
\end{equation}
Note that there is no forcing in $\dot{d}$ in that equation as it would be subdominant to that introduced by the impermeability condition, Eq.~\eqref{eq:imp_lub}.

At leading order in $\varepsilon$, the equations of motion for the fluid flow within the lubricating film are written in non-dimensional form as
\begin{align}
\frac{1}{r}\pard{(ru_r)}{r}+\pard{u_z}{z}=0,\qquad 
\pard{p}{r}=\pard{^2u_r}{z^2},\qquad
\pard{p}{z}=0.\label{eq:lub}
\end{align}
In the following, we focus exclusively on the leading-order contributions  remembering that higher order corrections  are at least $O(\varepsilon)$ smaller.

\subsection{Flow profile and pressure distribution}
 Solving the  system above follows  the classical approach~\citep{leal}. The dynamic pressure $p$ is invariant across the lubricating film and the previous equations can be integrated explicitly as
\begin{equation}
u_r=\frac{z^2}{2}\pard{p}{r}+\alpha z,\quad u_z=-\frac{z^3}{6 r}\pard{}{r}\left(r\pard{p}{r}\right)-\frac{z^2}{2 r}\pard{(r\alpha)}{r},
\end{equation}
where $\alpha(r,t)$ is determined by the dynamic boundary condition, Eq.~\eqref{eq:dynbclub},
\begin{equation}
\alpha=-\frac{h(h+2\lambda)}{2(h+\lambda)}\pard{p}{r}+\frac{\dot{a}}{h+\lambda}\pard{h}{r}\cdot\label{eq:alpha}
\end{equation}
From Eqs.~\eqref{eq:imp_lub}, \eqref{eq:h_lub} and \eqref{eq:alpha}, the Reynolds equation for the pressure gradient is obtained  as
\begin{equation}
\pard{h}{t}=\frac{1}{12r}\pard{}{r}\left[rh^3\left(\frac{h+4\lambda}{h+\lambda}\right)\pard{p}{r}-\frac{6rh^2\dot{a}}{h+\lambda}\pard{h}{r}\right]=\dot{d}-\frac{(h-d)\dot{a}}{a}\cdot
\end{equation}
This equation can be integrated in $r$ (since ${\partial p}/{\partial r}=0$ on the axis by symmetry), and using ${\partial h}/{\partial r}=r/a+O(\varepsilon^{3/2})$,  we obtain  
\begin{equation}
\pard{p}{r}=\left[\frac{6a(h+\lambda)\dot{d}}{h^3(h+4\lambda)}+\frac{3\dot{a}[h^2+(d-\lambda)h+\lambda d]}{h^3(h+4\lambda)}\right]\pard{h}{r}\cdot
\end{equation}
Integrating once more in $r$, the dominant pressure profile is obtained within the thin lubricating film as
\begin{align}
p(r)=p_\infty&+\frac{3\dot{a}}{64\lambda^2}\left[\left(3d-20\lambda\right)\log\left(1+\frac{4\lambda}{h}\right)-\frac{8d\lambda^2}{h^2}+\frac{4\lambda(4\lambda-3d)}{h}\right]+\frac{3a\dot{d}}{32\lambda^2}\left[3\log\left(1+\frac{4\lambda}{h}\right)-\frac{8\lambda^2}{h^2}-\frac{12\lambda}{h}\right],
\end{align}
where the radial dependence is implicit through the dependence on $h(r,t)$ and where $p_\infty$ is an $O(1)$ function of time only related to the pressure outside the lubricating film. In particular, in the two limit cases of a rigid shell or clean bubble, this profile simplifies as
\begin{align}
p(r)&=p_\infty-\frac{3a\dot{d}}{h^2}-\frac{3(d+2h)\dot{a}}{2h^2},\qquad \lambda=0 \textrm{   (rigid shell),}\label{eq:pressure_rigid}\\
p(r)&=p_\infty-\frac{3a\dot{d}}{4h^2}+\frac{3(2h-d)\dot{a}}{8h^2},\qquad \lambda=\infty \textrm{   (clean bubble).}\label{eq:pressure_bubble}
\end{align}

These results confirm the qualitative analysis of the flow field seen in figure~\ref{fig:flow_macro}. When $\lambda=0$, the pressure associated with the bubble's growth ($\dot{a}$) is lower than the background pressure, which results in an opposite forcing to that associated with the motion of the bubble's surface, and as a result the flow direction reverses across the lubrication film, revealing the existence of a recirculation region. In contrast, for $\lambda=\infty$, the pressure  associated with bubble growth is greater than the background pressure, forcing the flow in the same direction as the bubble's surface and thus the flow within the gap is oriented outward across the whole film.

\subsection{Forces on the inflating and translating bubble}
The dominant contribution to the vertical force on the bubble from the lubrication film is classically given by the pressure  and can be evaluated by integrating the pressure force over the region $0\leq r\leq R$ with $\sqrt{ad}\ll R\ll a$, leading to
\begin{align}
F_\textrm{lub}=&\Bigg[\frac{3\pi a\dot{a}}{8\lambda}\left\{(3d-20\lambda)\left[\left(1+\frac{h}{4\lambda}\right)\log\left(1+\frac{4\lambda}{h}\right)-1\right]+\frac{2d\lambda}{h}-16\lambda\log h\right\}\nonumber\\
&+\frac{3\pi a^2\dot{d}}{4\lambda}\left\{3\left[\left(1+\frac{h}{4\lambda}\right)\log\left(1+\frac{4\lambda}{h}\right)-1\right]+\frac{2\lambda}{h}\right\}\Bigg]_d^{h(R)}+O(1).
\end{align}
The dominant terms   arise from the central region ($r=0$ and $h=d$). Keeping in mind that we so far neglected corrections that are $O(\varepsilon)$ smaller, and that the matching with the outer solution introduces corrections that are $O(\varepsilon\log\varepsilon)$ smaller than the leading-order term as well as $O(1)$ pressure forces~\citep{cooley1969,kimbook}, the force coefficients can be rewritten
\begin{align}
C_T(\varepsilon,\tilde\lambda)&=-\frac{3\pi}{2\varepsilon}-\frac{9\pi}{4\tilde\lambda}\left[\left(1+\frac{\varepsilon}{4\tilde\lambda}\right)\log\left(1+\frac{4\tilde\lambda}{\varepsilon}\right)-1\right]+O(\log\varepsilon),\label{eq:force_coeffs1}\\
C_I(\varepsilon,\tilde\lambda)&=6\pi\log\varepsilon-\frac{3\pi}{8\tilde\lambda}(3\varepsilon-20\tilde\lambda)\left[\left(1+\frac{\varepsilon}{4\tilde\lambda}\right)\log\left(1+\frac{4\tilde\lambda}{\varepsilon}\right)-1\right]+O(1),\label{eq:force_coeffs2}
\end{align}
where we recall   that $\tilde\lambda=\lambda/a$ is the reduced slip length.

\renewcommand{\labelitemi}{$-$}
For arbitrary slip length, two different regimes can be identified depending on the relative magnitude of the three length scales of the problem (i.e.~$d$, $a$ and $\lambda$). When $\lambda\ll d\ll a$,  the dominant expression for the force coefficients is 
\begin{equation}
C_T(\varepsilon)=-\frac{6\pi}{\varepsilon}+O(\log\varepsilon),\qquad C_I(\varepsilon)=6\pi\log\varepsilon+O(1).\label{eq:force_coeff_sphere}
\end{equation}
This is essentially the result obtained from Eq.~\eqref{eq:pressure_rigid} in the case of a rigid sphere ($\lambda=0$). In contrast when $d\ll (a,\lambda)$, the dominant expression for the force coefficients is now
\begin{equation}
C_T(\varepsilon)=-\frac{3\pi}{2\varepsilon}+O(\log\varepsilon),\qquad C_I(\varepsilon)=-\frac{3\pi\log\varepsilon}{2}+O(1),\label{eq:force_coeff_bubble}
\end{equation}
which  is   the result obtained from Eq.~\eqref{eq:pressure_bubble} in the case of a \sm{perfectly-clean} bubble ($\lambda=\infty$). Note that the scaling for the leading order correction in Eqs.~\eqref{eq:force_coeffs1}--\eqref{eq:force_coeff_bubble} is only shown with respect to $\varepsilon$ for fixed $\tilde\lambda$, and may include singular terms when $\tilde\lambda\ll 1$ (see below).

Noticeably, for a sufficiently small gap, the leading-order values of the force coefficients do not depend on the particular value chosen for $\tilde\lambda$. Two regimes can therefore be distinguished for gap width smaller than the slip length (\sm{clean} bubble limit) or larger than the slip length (rigid boundary limit) and the calculation above suggests a transition between the two regimes for $d\sim \lambda$.

\begin{figure}
\begin{center}
\begin{tabular}{c}
\includegraphics[width=.7\textwidth]{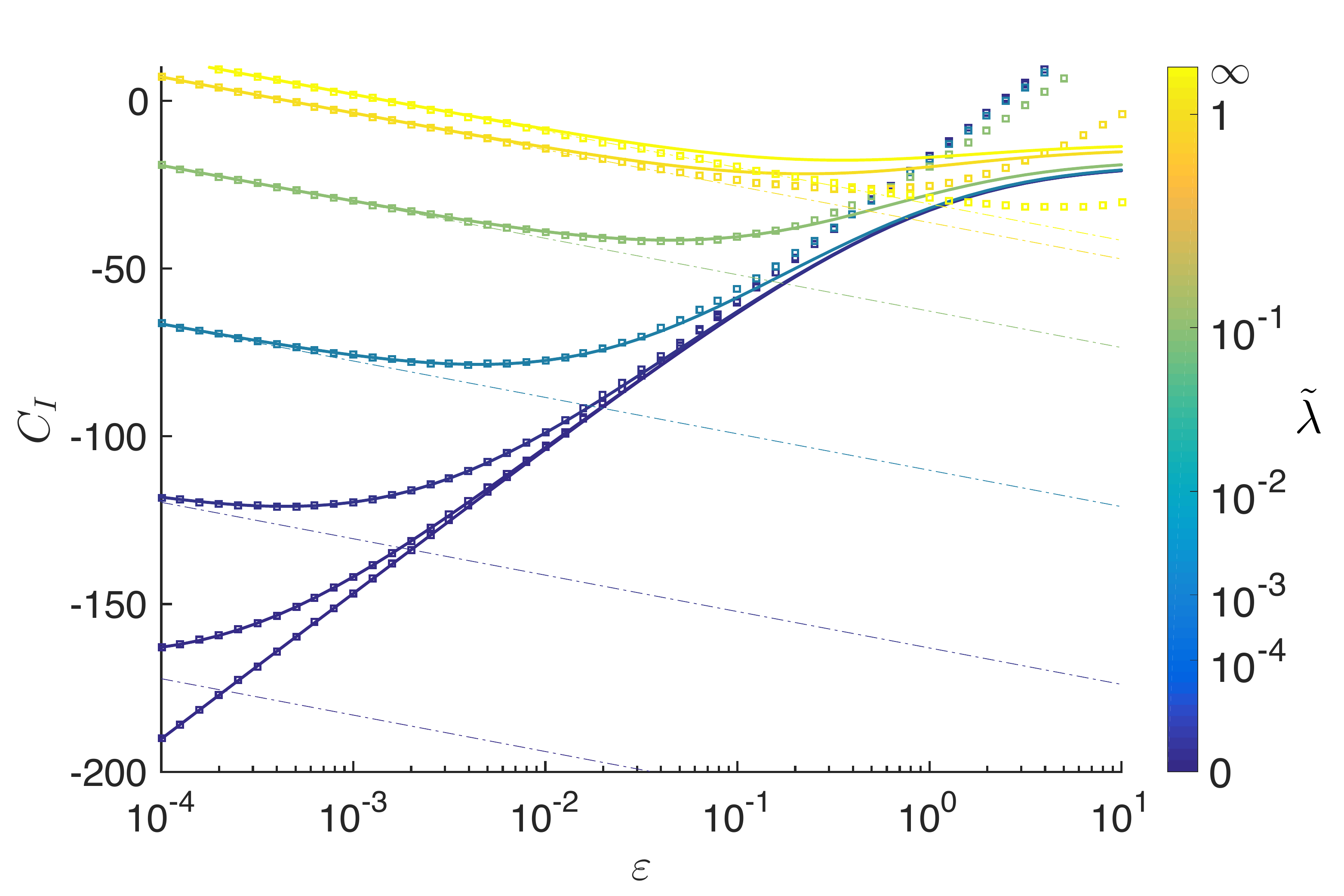}\\
\includegraphics[width=.7\textwidth]{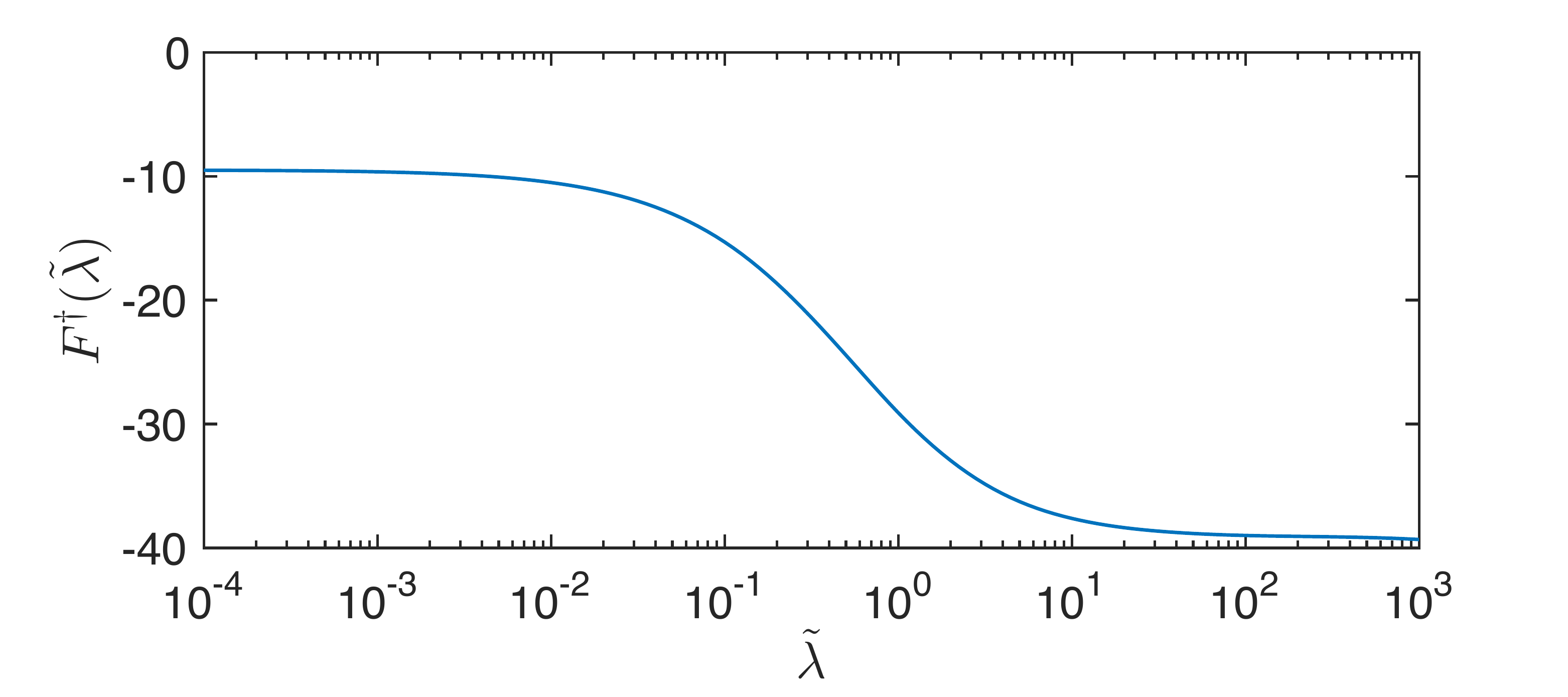}\\
\includegraphics[width=.7\textwidth]{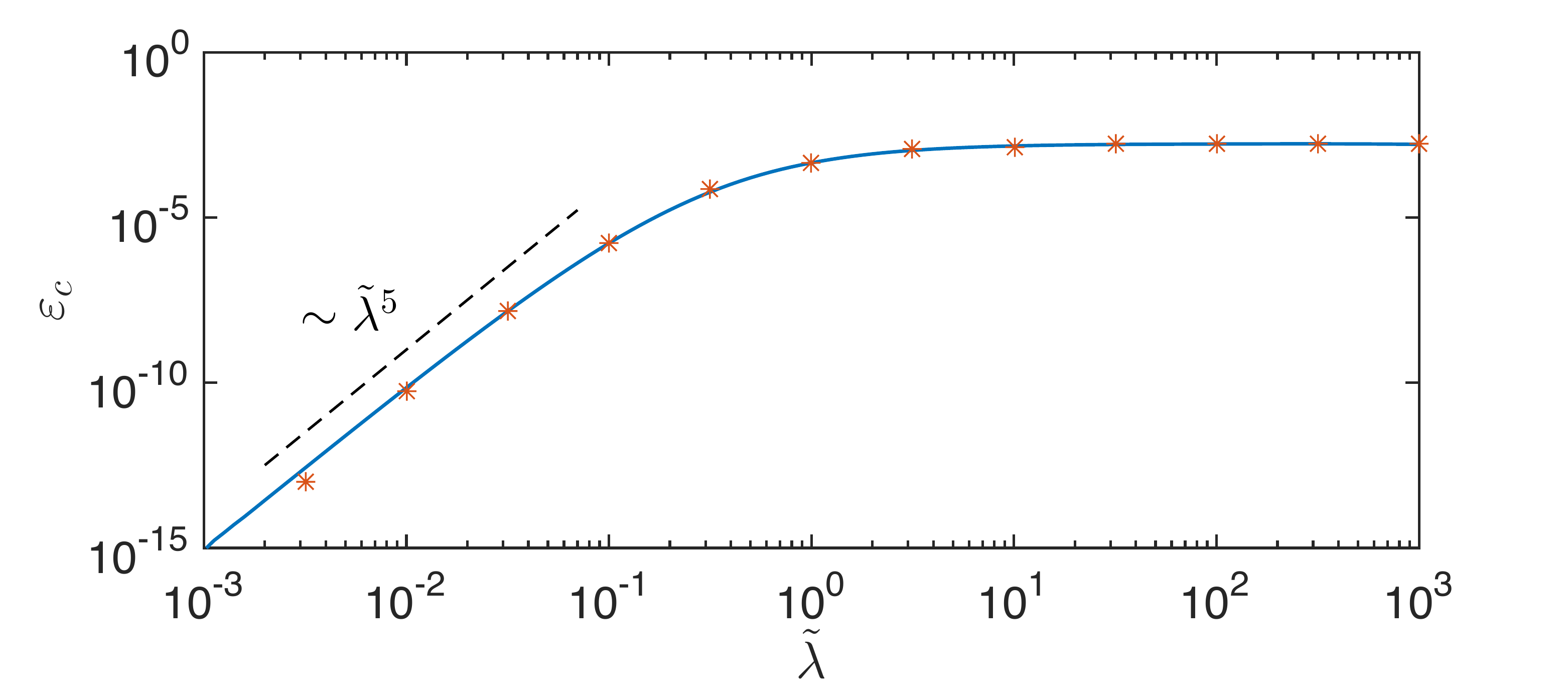}
\end{tabular}
\caption{Top: Dependence of the force coefficient associated with bubble growth (for fixed gap width) with the relative gap width, $\varepsilon$, and different values of the relative slip length, $\tilde\lambda$. For each value of $\tilde\lambda$, the full semi-analytical solution from Section~\ref{sec:model} is shown (solid line) and compared with the asymptotic approximations $C_I^*$ in Eq.~\eqref{eq:CIapprox} (dash-dotted line) and $C_I^\dag$ in Eq.~\eqref{eq:CIapprox2} (square symbols). 
Centre:  Function  $F^\dag(\lambda)$ as defined in Eq.~\eqref{eq:CIapprox2}. 
Bottom: Dependence of $\varepsilon_c$ on $\tilde\lambda$, such that $C_I>0$ for $\varepsilon\leq\varepsilon_c$. The results obtained from Eq.~\eqref{eq:CIapprox2} are shown as a solid line while the symbols correspond to the full semi-analytical solution from Section~\ref{sec:model}.}
\label{fig:CIapprox}
\end{center}
\end{figure}

For all $\lambda\neq 0$, an asymptotic approximation of $C_I$ can therefore be obtained as
\begin{equation}
C_I^*(\varepsilon,\tilde\lambda)=-\frac{3\pi}{2}\log\varepsilon+F^*(\tilde\lambda),\label{eq:CIapprox}
\end{equation}
where $F^*(\tilde\lambda)$ is independent of $\varepsilon$, and defined such $C_I^*(\varepsilon^*,\tilde\lambda)=C_I(\varepsilon,\tilde\lambda)$ for a given value of $\varepsilon^*\ll 1$ (note that the value of $F(\tilde\lambda)$ obtained   does not depend on $\varepsilon^*$ provided it is small enough; $\varepsilon^*=10^{-5}$ was used here). However, the dominant term in $C_I$ is only $O(\log\varepsilon)$ and for finite $\tilde\lambda\ll 1$, a very small gap $\varepsilon$ might be required to obtain a correct approximation of the force due to bubble growth using Eq.~\eqref{eq:CIapprox}. This asymptotic result for 
the hydrodynamic force on the bubble agrees very well with the full semi-analytical solution from Section~\ref{sec:model}, as illustrated in figure~\ref{fig:CIapprox} (top).

Equation~\eqref{eq:force_coeffs2} may yet provide a significantly improved estimation of the force coefficient for small but arbitrary slip length. Specifically,  isolating singular terms in $\log\tilde\lambda$ introduced for $\varepsilon\ll 1$ by the form of Eq.~\eqref{eq:force_coeffs2} leads to
\begin{equation}
C_I^\dag(\varepsilon,\tilde\lambda)=6\pi\log\varepsilon-\frac{3\pi}{8\tilde\lambda}(3\varepsilon-20\lambda)\left[\left(1+\frac{\varepsilon}{4\tilde\lambda}\right)\log\left(1+\frac{4\tilde\lambda}{\varepsilon}\right)-1\right]-\frac{15\pi}{2}\log(1+4\tilde\lambda)+F^\dag(\tilde\lambda)\label{eq:CIapprox2},
\end{equation}
where $F^\dag(\tilde\lambda)$ is defined using the same approach as $F^*$ above. Physically, $F^\dag(\tilde\lambda)$ includes contribution of the outer problem (the hydrodynamic force introduced on the top of the sphere for example) as well as $O(1)$ contributions already included in $C_I^\dag$. Figure~\ref{fig:CIapprox} demonstrates that this approch provides a much improved approximation of the force induced by the bubble growth dynamics (including the transition between the two regimes $\lambda=0$ and $\lambda=\infty$ for small gaps). A more quantitative comparison would actually show that $|C_I-C_I^\dag|$ is $O(\varepsilon)$ for all $\lambda$ when $\varepsilon\rightarrow 0$. Also it should be noted that $F^\dag(\tilde\lambda)$ remains finite for all values of $\tilde\lambda$ and in particular in the limit $\tilde\lambda\ll 1 $ and $\tilde\lambda\gg 1$ (Figure~\ref{fig:CIapprox}, middle).

\subsection{Emergence of a repulsive force for a growing bubble with fixed gap}
\label{sec:force_rev}
These results allow for a quantitatively accurate estimate of the critical aspect ratio $\varepsilon_c(\tilde\lambda)$ that leads to $C_I>0$  and the onset of repulsion from the wall induced by bubble growth (for a fixed gap). Such sign change always occurs when $\varepsilon\ll \tilde\lambda$, i.e.~when the gap width is much smaller than the slip length (Figure~\ref{fig:CIapprox}). For large $\tilde\lambda$, $\varepsilon_c$ converges to a finite value $\varepsilon_c^{\tilde\lambda=\infty}\approx 1.8\:10^{-3}$  (Figure~\ref{fig:CIapprox}, bottom).  When $\varepsilon\ll\tilde\lambda\ll 1$,  Eq.~\eqref{eq:force_coeffs2}  takes the form
\begin{equation}
C_I=-\frac{3\pi}{2}\log\left(\frac{\varepsilon}{\tilde\lambda^5}\right)+O(1)\label{eq:CI_final},
\end{equation} 
where the $O(1)$ term here remains finite when $\tilde\lambda\rightarrow 0$ and $\varepsilon/\tilde\lambda\rightarrow 0$. The critical aspect ratio $\varepsilon_c(\tilde\lambda)$ corresponds to the value of $\varepsilon$ leading to $C_I=0$. When $\tilde\lambda\rightarrow 0$, $\varepsilon_c(\tilde\lambda)$ must therefore scale as $\varepsilon_c\sim\tilde\lambda^5$ which is  consistent with the results of the exact semi-analytical solution, as shown in figure~\ref{fig:CIapprox} (bottom).

\section{Scalings for the growth of a free bubble}
\label{sec:consequences}
Our analysis   of the instantaneous fluid flow within the lubrication gap and associated forces allows for a better understanding of the non-monotonic dynamics reported in Section~\ref{sec:growth} for the growth of a force-free bubble, as well as the associated asymptotic behaviour of the film thickness. In Section~\ref{sec:growth}, three different regimes were identified: a linear decay of the gap width with bubble radius when the bubble is far from the wall ($d\gg a$, regime A), a linear growth of the lubrication film with bubble radius when the relative gap width is small and the relative slip length is large ($d\ll a\ll \lambda$, regime B) and an exponential decay of the lubrication film thickness with bubble radius when the slip length becomes smaller than the bubble radius ($d\ll a$ and $\lambda\ll a$, regime C). The lubrication analysis above is only applicable to  regimes B and C.

\subsection{Asymptotic behaviour for regime B}
When $d\ll a\ll \lambda$, Eq.~\eqref{eq:force_coeff_bubble} provides the leading-order dynamics of the film as
\begin{equation}
\totd{D}{a}=-\frac{C_I(D/a,\lambda/a)}{C_T(D/a,\lambda/a)}=-\frac{D}{a}\log\left(\frac{D}{a}\right),
\end{equation}
whose solution for some initial (or intermediate) condition $D(a^*)=d^*$ is obtained as
\begin{equation}
D(a)=a\ee\left(\frac{d^*}{\ee a^*}\right)^{a^*/a}.
\end{equation}
As $a\gg a^*$, the film thickness grows linearly with $a$, and $\varepsilon$ remains roughly constant, consistently with the results presented on figures~\ref{fig:growth_bubble} and~\ref{fig:growth_general} (regime B, bottom left). This regime, and the approximation underlying Eq.~\eqref{eq:force_coeff_bubble}, eventually breaks down when $a\sim\lambda$ (transition to regime C).

\subsection{Asymptotic behaviour for regime C}
When $\lambda\ll d\ll a$, Eq.~\eqref{eq:force_coeff_sphere} provides the leading-order dynamics of the film as
\begin{equation}
\totd{D}{a}=-\frac{C_I(D/a,\lambda/a)}{C_T(D/a,\lambda/a)}=\frac{D}{a}\log\left(\frac{D}{a}\right),
\end{equation}
whose solution for $D(a^*)=d^*$ is
\begin{equation}
D(a)=a\ee\left(\frac{d^*}{\ee a^*}\right)^{a/a^*},
\end{equation}
and decays exponentially with $a$, consistently with figure~\ref{fig:growth_general} (regime C). Note however, that this regime cannot describe the final dynamics of the film, since within this regime $\varepsilon$ decays exponentially, but $\tilde\lambda$ only decays algebraically. Eventually, $d$ becomes (much) smaller than $\lambda$, which violates the assumptions for regime C. 

\subsection{Final dynamics}
The final dynamics correspond to a very large bubble radius compared to both $d_0$ and $\lambda$. This final regime satisfies therefore necessarily $(\varepsilon,\tilde\lambda)\ll 1$. The previous section showed that $\tilde\lambda\ll \varepsilon$ cannot hold indefinitely, so in the final regime $\varepsilon$ is at most $O(\tilde\lambda)$. Maintaining $\varepsilon=O(\tilde\lambda)$ asymptotically is however impossible since $\varepsilon>\varepsilon_c\sim\tilde\lambda$ leads to a decrease in the lubrication film thickness $d$ with $a$ so that $\varepsilon/\tilde\lambda\ll 1$ eventually.

The only possible configuration for the final dynamics is therefore $\varepsilon\ll \tilde\lambda\ll 1$. Then, the dominant scaling of Eq.~\eqref{eq:CI_final} applies and the leading-order dynamics is obtained as
\begin{equation}
\totd{D}{a}=-\frac{C_I(D/a,\lambda/a)}{C_T(D/a,\lambda/a)}=-\frac{D}{a}\log\left(\frac{Da^4}{\lambda^5}\right),
\end{equation}
whose solution for $D(a^*)=d^*$ is given by
\begin{equation}
D(a)=\frac{\lambda^5\ee^4}{a^4}\left(\frac{d^*a^{*4}}{\lambda^5\ee^4}\right)^{a^*/a}.
\end{equation}
The final dynamics of the film follows $\varepsilon\sim\tilde\lambda^5$ and thus amounts to an algebraic decay of the lubrication film thickness as $d\sim \lambda^5/a^4$. For force-free growing bubbles, this   corresponds to the accumulation of trajectories at the lower boundary of regime C in figure~\ref{fig:growth_general}.  This final algebraic regime is therefore universal: regardless of the value of $\lambda$ (provided it is not strictly zero or infinite), the lubrication film thickness always eventually decays algebraically with the bubble radius and reaches contact in asymptotically infinite time, whether after a transient growth of the film thickness (for $\lambda\gtrsim d_0$) or after a monotonic and transient exponential decay ($\lambda\lesssim d_0$).

\section{Discussion}
\label{sec:conclusions}
The results presented in this paper provide fundamental insight into the viscous hydrodynamic interaction of a force-free growing bubble with a nearby rigid wall and on the influence of the bubble surface properties. These are  represented here as a simple Navier slip length $\lambda$, thus accounting for various surface behaviours  ranging from perfectly-clean stress-free surfaces ($\lambda=\infty$) to surfactant-laden slip-free surfaces ($\lambda=0$). An analytical solution for the flow field and bubble motion was first obtained for arbitrary bubble size and position, providing a complete description of the fluid dynamics and hydrodynamic forces. Asymptotic analysis of the fluid motion within the fluid gap using lubrication theory shed further light on the  forces associated with the bubble growth, their fundamental scalings and the structure of the flow within the gap.

In contrast to classical problems on lubrication film dynamics such as the sedimentation of a rigid sphere toward a wall, the drainage of the film is forced here by the growth of the bubble, which squeezes fluid out of the interstitial fluid layer. Neglecting buoyancy effects, the net hydrodynamic force on the bubble must vanish and viscous  drag must therefore balance  the forces arising from the growth of a fixed bubble so that a net displacement of the bubble's centre is  in general  necessary. While intuition suggests that the fluid layer between the bubble and the wall must be drained through this process, non-trivial variations of its thickness were revealed in our analysis, varying critically with the surface properties of the  bubble. 

Starting from a bubble of infinitesimal size located at a finite distance $d_0$ from the wall, the fluid motion within the interstitial gap is observed to follow two different routes depending on the relative ratio of the initial distance $d_0$ to the slip length $\lambda$ of the  bubble surface. While both routes share the same initial and final features they differ in their transient dynamics. In all cases, the initial dynamics correspond to a linear closing of the fluid gap when the bubble is much smaller than the initial distance, as its centre remains fixed. Since the wall effect is negligible, the radial flow resulting from bubble growth does not generate any net force leading to no significant motion of the bubble centroid. 

Confinement effects become significant when the bubble radius and distance to the wall become comparable. For relatively small slip length ($\lambda\ll d_0$), confinement-induced translation of the bubble's centre remains slightly slower than the bubble's growth rate, resulting in the monotonic and exponential decrease of the thickness of the lubricating film. In contrast, bubbles with relatively large slip length ($\lambda\gg d_0$) experience a rebound in the film thickness once it is much smaller than the bubble radius (i.e.~confinement-induced translation of their centre of mass is a bit faster than bubble growth). During that rebound, the minimum fluid layer thickness grows almost linearly with the bubble radius. This effective growth-induced repulsion stops when the bubble radius becomes of the same order as the slip length, at which point the thickness of the fluid layer  starts decreasing again. This second transition occurs when the resistance to bubble growth applied by the outer fluid becomes larger than the repulsion introduced by the lubricating film. These two different behaviours and scalings are  linked to the fluid flow and pressure distribution within the lubrication film, which originate in the different ability of stress-free and no-slip surfaces to drive flow within the fluid gap.  For all values of $\lambda$, the thickness of the interstitial fluid layer eventually decays algebraically as the inverse fourth power of the bubble radius, and this scaling is maintained until contact with the wall which is achieved only asymptotically for infinite time in the absence of any surface roughness or short-range bubble-wall interaction. 

\sm{The analysis presented in the paper focuses on purely viscous flows with no external forces. Inertial and transient effects are negligible while the Reynolds number \mbox{Re}, which scales with the radius of the bubble, remains small, i.e. $a^*\ll \eta/\rho U^*$, with $U^*$ being the characteristic growth speed  of the bubble radius. Similarly, gravitational forces are negligible for small bubbles (they scale with the cube of the bubble radius) and remain so provided that $a^*\ll \sqrt{\eta U^*/\rho g}$. Both conditions are satisfied for microscopic bubbles and during the early stages of the growth of larger bubbles. The previous two conditions further provide a limit on the bubble size beyond which the present model is no longer valid.}

The modelling approach in the paper focuses \sm{also} exclusively on non-deformable bubbles (i.e.~small capillary numbers). This  amounts to a quasi-steady assumption regarding the growth dynamics of the bubble which is prescribed here as an input forcing. As a result, because of the fundamental properties of Stokes' equations, the problem is fully reversible and all the conclusions drawn here could be used also to describe the reverse process of bubble dissolution. Also, the phase-space trajectories of the system (i.e.~the variations of the film thickness with bubble radius) do not depend on the rate at which these trajectories are followed. In the case of fast bubble growth, hydrodynamic stresses (in particular lubrication forces) may become dominant over their capillary counterpart, leading to significant bubble deformations in response to confinement. An important extension of the present problem would therefore account for bubble deformation. In order to obtain the fluid film and bubble dynamics, and in contrast with the present situation where the bubble geometry is defined by a single degree of freedom (its radius), such an approach would require obtaining the dynamic shape of the bubble by enforcing the mechanical balance at each point of the bubble surface~\citep{bretherton1961}, a problem related  to that of the interaction of colliding soft spheres~\citep*{davis1986}.

Throughout this paper, the bounding surface was assumed to be a no-slip wall. The results presented here demonstrate that the hydrodynamic nature of the bubble's surface plays a critical role in determining the transient drainage dynamics of the fluid film, and a natural extension of this work should indeed consider how the present results are modified in the vicinity of a free surface (no-stress) or more generally a surface characterised by a second slip length $\lambda^*$. The case of a free surface ($\lambda^*=\infty$) is expected to lead to significantly different dynamics, as it allows for a fundamental reduction of the lubricating fluid resistance to the approach of the spherical bubble. Classical results have indeed established that the normal force on a droplet translating toward a free surface scales as $\sqrt{a/d}$ (instead of $a/d$ for a rigid wall)~\citep{barnocky1989}. The case of a free surface would also allow to  model mathematically  the synchronised growth of two identical bubbles since in that case the plane of symmetry separating the two bubbles is a no-shear surface.

Admittedly, the most severe modelling assumption of our work is the use of a  constant, uniform Navier slip length model     to account for the variability in the state of the bubble surface and its effective physico-chemical properties. This model is  too simplistic to describe complex and time-dependent surface processes  such as surfactant adsorption-desorption or reorganisation on a liquid-gas interface. In particular, it ignores variations of the effective slip length (and surface tension) with local bubble curvature and the  coupling between the hydrodynamic problem and the local surface behaviour. Yet, the simplicity of this model, and its relative generality, allow to illustrate the singular nature of clean bubbles
and provide  valuable physical insight into the interplay between hydrodynamics and surface mobility. 
\\

This project has received funding from the European Research Council (ERC) under the European Union's Horizon 2020 research and innovation programme under Grant Agreements 714027 (SM), 280117 (FG) and 682754 (EL). This project also received financial support from the Swiss National Science Foundation (SNFS) with the Doc.Mobility Fellowship P1ELP2-172277.

\end{document}